\documentclass[10pt,
  DIV=12,
  parskip=half]{scrartcl}
\usepackage{xcolor}
\usepackage{amssymb}

\usepackage{microtype}
\usepackage{physics, units, dsfont, bm}
\usepackage{mathtools, amsfonts, mathrsfs, bbm}
\usepackage{graphicx}
\usepackage{booktabs}
\usepackage{fancyvrb}
\usepackage{lipsum}

\usepackage{amsthm}
\setkeys{Gin}{width=\linewidth,totalheight=\textheight,keepaspectratio}
\graphicspath{{graphics/}}
\fvset{fontsize=\normalsize}

\newcommand{\kgfigurefadeopacity}{0.07}

\definecolor{rememberblue}{RGB}{28,92,140}
\definecolor{rememberlight}{RGB}{238,247,252}
\definecolor{rememberborder}{RGB}{120,170,200}

\newsavebox{\rememberboxcontent}

\definecolor{openpurple}{RGB}{96,62,142}
\definecolor{openlight}{RGB}{248,245,252}
\definecolor{openborder}{RGB}{165,135,205}

\newsavebox{\openquestionboxcontent}

\usepackage{tikz,tikz-3dplot}
\usetikzlibrary{fit,positioning}
\usetikzlibrary{arrows.meta}
\usetikzlibrary{calc}
\usepackage{todonotes}
\usepackage{float}
\newtheorem{theorem}{Theorem}
\numberwithin{theorem}{section}
\newtheorem{lemma}[theorem]{Lemma}

\theoremstyle{definition}
\newtheorem{definition}[theorem]{Definition}
\newtheorem{property}[theorem]{Property}

\newtheorem{example}[theorem]{Example}

\newcommand{\graphNoEdge}{%
  \begin{tikzpicture}[baseline=-0.5ex, scale=0.3]
    \tikzstyle{every node}=[circle,fill,inner sep=0pt,minimum size=3pt]
    \node (a) at (0,0) {};
    \node (b) at (1,0) {};
  \end{tikzpicture}%
}

\newcommand{\graphEdge}{%
  \begin{tikzpicture}[baseline=-0.5ex, scale=0.3]
    \tikzstyle{every node}=[circle,fill,inner sep=0pt,minimum size=3pt]
    \node (a) at (0,0) {};
    \node (b) at (1,0) {};
    \draw (a) -- (b);
  \end{tikzpicture}%
}

\makeatletter
\newcommand*{\centerfloat}{%
  \parindent \z@
  \leftskip \z@ \@plus 1fil \@minus \textwidth
  \rightskip\leftskip
  \parfillskip \z@skip}
\makeatother

\tikzset{
    kgfigedge/.style={
        line width=1.8pt,
        line cap=round,
        line join=round,
        preaction={draw=white, line width=6pt}
    },
    kgfigfadededge/.style={
        kgfigedge,
        opacity=\kgfigurefadeopacity
    },
    kgfigqone/.style={kgfigedge, draw=red},
    kgfigqtwo/.style={kgfigedge, draw=blue},
    kgfigqthree/.style={kgfigedge, draw=green!60!black},
    kgfignode/.style={
        circle,
        draw,
        fill=white,
        line width=1.6pt,
        minimum size=9.2mm,
        inner sep=0pt,
        font=\scriptsize
    },
    kgfigsmallnode/.style={
        circle,
        draw,
        fill=white,
        line width=1.35pt,
        minimum size=7.5mm,
        inner sep=0pt,
        font=\scriptsize
    },
    kgfigbox/.style={
        rounded corners=4pt,
        draw=black!55,
        fill=white,
        line width=1.2pt,
        inner xsep=10pt,
        inner ysep=8pt
    },
    kgfiglabel/.style={font=\scriptsize},
    kgfigarrow/.style={kgfigedge, -{Latex[length=2.6mm,width=1.8mm]}}
}

\usepackage{pifont}
\usepackage{hyperref}
\usepackage{cleveref}

\begin{document}

\title{What are local complementations, really?}
\author{%
    Kenneth Goodenough\thanks{Email: kdgoodenough@gmail.com}\\
 {\footnotesize Naturwissenschaftlich-Technische Fakultät, Universität Siegen, Walter-Flex-Straße 3, 57068 Siegen, Germany}}
\date{}

\maketitle

\vspace{-10mm}
\begin{abstract}
A local complementation at a vertex $v$ of a graph $G$ replaces the induced neighborhood of $v$ with its complement. Local complementations are, at face value, peculiar operations. Yet they famously show up when describing the local Clifford equivalence among graph states, and, because of this, have become a fundamental notion in the study of graph states. Following Bouchet, we show here how local complementations on graphs are more naturally thought of in terms of \emph{switchings} of \emph{(matroid) bases of stabilizer states}. Bases and switchings reveal structural properties of stabilizer states that are harder to glean from the graph-theoretic description alone, such as vertex-minors and statistics of local Pauli measurements. Bases also simplify prior proofs about graphs up to local complementations, of which we give several examples.
Bases are foundational in the study of (multi)matroids; as such this paper serves as a gentle introduction to the applications of multimatroids to quantum information theory.
\end{abstract}

\section{Introduction}
\subsection{Motivation}
A local complementation at a vertex $v$ replaces the induced neighborhood of that vertex with its complement~\cite{fon1996local}. Famously, two graph states $\ket{G}$ and $\ket{H}$ are related by local Cliffords if and only if $G$ and $H$ are equivalent up to a sequence of local complementations~\cite{van2004graphical}. This connection has motivated substantial work in quantum information on studying stabilizer states through graphs modulo local complementation~\cite{dahlberg2018transforming, goodenough2024bipartite, goodenough2024near, chitambar2025quantum, bhatti2025distributing, sharma2025minimizing, dahlberg2020counting, dahlberg2022complexity, adcock2020mapping, claudet2024local, claudet202627, claudet2025deciding, hahn2026structure, cattaneo2012mindegree, hoyer2006resources, burchardt2025foliage, hahn2019quantum, hahn2022limitations, de2024extracting}.

Characteristically, local complementations were studied by mathematicians decades before they made their way to the quantum world. One of the first mathematicians to do so was Kotzig~\cite{kotzig68}, followed by Rosenstiehl, Read~\cite{rosenstiehl1976gauss}, and de Fraysseix~\cite{de1981local}, showing surprisingly deep connections with other topics in combinatorics. These ideas were taken up by Bouchet~\cite{bouchet1988transforming, bouchet1993recognizing}, who soon realized the connection between graphs up to local complementations, $4$-regular multigraphs~\cite{bouchet1994circle, bouchet1996connectivity}, and isotropic systems~\cite{bouchet1987isotropic, bouchet1989connectivity, bouchet1988graphic}, the latter of which are known to be stabilizer groups when ignoring phases of the stabilizers~\cite{dahlberg2020counting}. Bouchet's line of thought culminated with his introduction of \emph{multimatroids}, a combinatorial object that vastly generalizes matrices and graphs, in an attempt to better understand local complementations~\cite{bouchet1997multimatroids, bouchet1998multimatroids, bouchet2001multimatroids, bouchet19984multimatroids}. More specifically, Bouchet found that graphs up to local complementations are better understood as \emph{(matroid) bases up to switchings}. This leads, among other things, to simpler proofs on stabilizer entanglement, of which we provide several examples. We note that some of these concepts have implicitly appeared already in the quantum literature, albeit without providing a physical interpretation of bases and switchings, see e.g. Refs.~\cite{dahlberg2018transforming, dahlberg2020transform, dahlberg2022complexity}. 

This paper provides an exposition of some of Bouchet's ideas in terms of quantum information, alongside some new results. We are hopeful that these ideas will provide a more general perspective on stabilizer entanglement than can be obtained by restricting attention to graph states alone \cite{goodenough2026asymptotic}.

\subsection{Overview}
The connection between graphs and bases can be summarized with the following diagram.
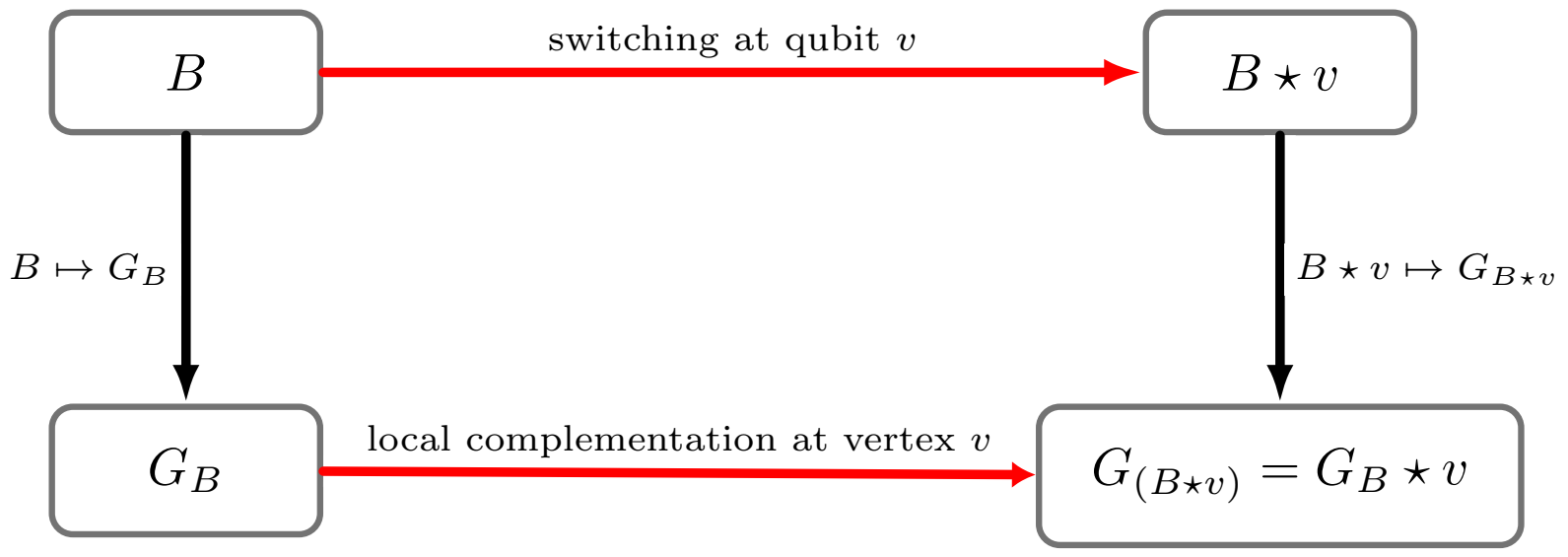
\vspace{-6mm}
\begin{figure}[H]
    \centerfloat
% Horizontal separation between left- and right-hand columns
\newcommand{\kghorizsep}{55mm}

\begin{figure}[H]
    \centering
    \begin{tikzpicture}[node distance=18mm and \kghorizsep]
        \node[kgfigbox, minimum width=18mm] (B) {$B$};
        \node[kgfigbox, minimum width=18mm, right=of B] (Bs) {$B\star v$};
        \node[kgfigbox, minimum width=18mm, below=of B] (GB) {$G_B$};
        \node[kgfigbox, minimum width=25mm, below=of Bs] (GBs) {$G_{(B\star v)}=G_{B}\star v$};

        \draw[kgfigqone,-{Latex[length=2.6mm,width=1.8mm]}]
            (B) -- node[above,inner sep=2pt,font=\scriptsize]{switching at qubit $v$} (Bs);
        \draw[kgfigarrow]
            (B) -- node[left,fill=white,inner sep=2pt,font=\scriptsize]{$B\mapsto G_B$} (GB);
        \draw[kgfigarrow]
            (Bs) -- node[right,fill=white,inner sep=2pt,font=\scriptsize]{$B\star v\mapsto G_{B\star v}$} (GBs);
        \draw[kgfigqone,-{Latex[length=1.6mm,width=1.8mm]}]
            (GB) -- node[above,inner sep=2pt,font=\scriptsize]{local complementation at vertex $v$} (GBs);

        % \node[kgfiglabel,below=3mm of GBs] {$G_{B\star v}=G_B\star v$};
    \end{tikzpicture}
\end{figure}
\end{figure}
\vspace{-9.5mm}
At the bottom we have the familiar picture: a graph $G_B$ is transformed to another graph $G_B\star v$ by locally complementing at the vertex $v$. At the top we have a \emph{basis} $B$ being mapped to a new basis $B\star v$ by \emph{switching} at qubit $v$; we will define bases and switchings later. For now, suffice to say that a basis of a stabilizer $\ket{\psi}$ is a full-weight Pauli string satisfying a certain property with respect to $\ket{\psi}$. Furthermore, for each qubit $v$, switching a basis $B$ at $v$ yields a new basis $B\star v$. Note the analogy with graphs and local complementations: for each $v$, locally complementing $G$ at $v$ yields a graph $G\star v$. 

This analogy is made precise by the map $B\mapsto G_B$; for each basis $B$, there is a unique graph---called the \emph{fundamental graph} $G_B$ with respect to $B$---such that $G_B\star v=G_{(B\star v)}$. That is, locally complementing a fundamental graph $G_B$ at $v$ yields the fundamental graph $G_{(B\star v)}$ with respect to the switched basis $B\star v$.

In section \ref{sec:fundamental_graphs} we will define and explain bases and fundamental graphs; we show how switchings on bases abstract local complementations in section \ref{sec:local_comps_as_switches}. In section \ref{sec:use_of_bases} we will discuss some of the benefits bases have over graphs. We will show that 1) bases allow for simplified proofs of statements of graphs up to local complementations, 2) show that the map $B\mapsto G_B$ is `lossy', and how this is connected with local symmetries of the underlying state, and 3) that bases are well-behaved under local Pauli measurements, and use this to relate bases to so-called \emph{vertex-minors}. We conclude in section \ref{sec:multimatroids} by discussing how the study of bases naturally leads one to study stabilizer states through the lens of multimatroids.

\subsection{Notation, conventions and terminology}\label{sec:notation}
We will assume basic familiarity with graph theory/graph states, linear algebra, and the stabilizer formalism. We'll freely convert between vertices and qubits depending on the context, and use $V=\lbrace 1, 2, \ldots, n \rbrace$ for the set of vertices/qubits. Adjacency of two vertices $v, w$ of a graph $G$ is denoted by $v \sim_G w$. The neighborhood of a vertex $v$ in a graph is denoted by $N_v$, and the closed neighborhood by $\overline{N}_v=N_v \cup \lbrace{v\rbrace}$. For a Pauli string $P$, $P_v$ denotes the entry of $P$ at $v$. The canonical generators of a graph state $\ket{G}$ are of the form $X_vZ_{N_v}$ for each qubit $v$. A local complementation at a vertex $v$ is denoted by $G\mapsto G\star v$, and we say that two graphs are locally equivalent if they are related by a sequence of local complementations. Following~\cite{goodenough2026asymptotic}, we extend this terminology to arbitrary stabilizer states: $\ket{\psi}$ and $\ket{\phi}$ are locally equivalent if the states are equivalent up to local Cliffords, and write this as $\ket{\psi}\simeq_{\textrm{LC}}\ket{\phi}$. 

We'll ignore phases of all Pauli strings, since it is in this setting that the connection with multimatroids/bases becomes most apparent. This also means that we treat $X$, $Y$, $Z$ as just labels, and not as matrices. For each $v$, the \emph{skew class} is $\omega_v=\lbrace X_v, Y_v, Z_v\rbrace$, the set of non-identity Paulis acting on $v$. Single-qubit Cliffords have then a simple interpretation: they are precisely the six ways of relabeling a skew class. Importantly, it will be convenient to treat Pauli strings $P$ not just as the tensor product of all the $\lbrace{P_v\rbrace}_{v\in V}$, but as sets containing the non-identity elements of $P$. As an example, $P=IXYZ \simeq \lbrace X_2, Y_3, Z_4\rbrace$, so that $IIYZ\subseteq P$, but $XXYI\nsubseteq P$. Finally, $A\Delta B$ denotes the symmetric difference of the sets $A$ and $B$.

\section{Fundamental graphs and bases}\label{sec:fundamental_graphs}
We first motivate fundamental graphs by showing how they recover the graphs of graph states. Afterwards we define bases and give them a physical interpretation, and show how bases can be used to define fundamental graphs for arbitrary stabilizer states.

\subsection{A simple graph state example}\label{sec:a_simple}
Assume we fix some graph state $\ket{G}$ on qubits $V$, with $|V|=n$. Let $B=Z^{\otimes n}$ be the all-$Z$ string. We'll use the graph state below as an example in this subsection.
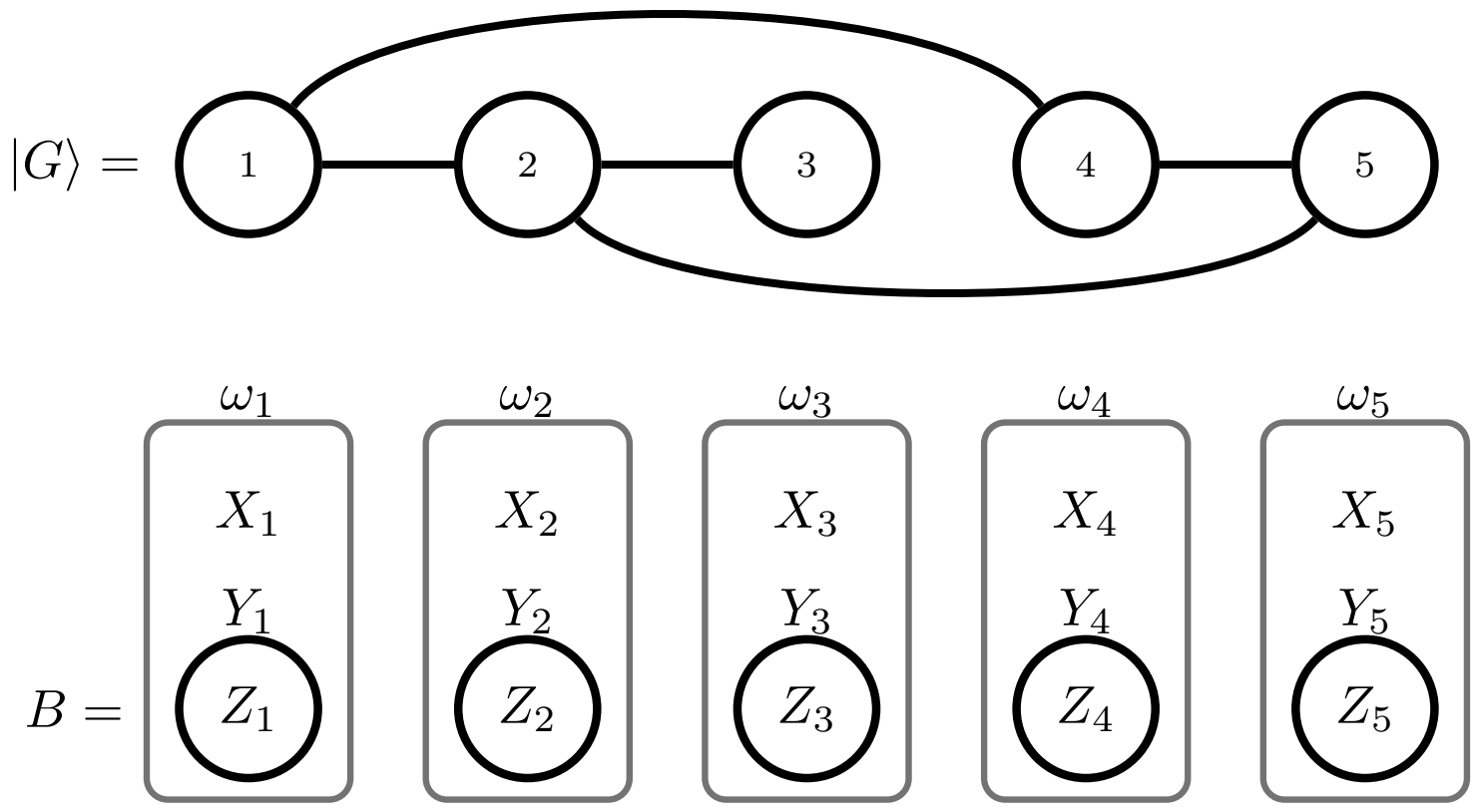
\vspace{-5mm}
\begin{figure}[H]
    \centerfloat
\begin{figure}[H]
    \centering
    \begin{tikzpicture}[scale=0.925]

        %------------------------------------------------
        % Graph: vertices aligned with the skew classes
        %------------------------------------------------
        \node at (-5.25,3) {$\ket{G}=$};

        \node[kgfignode] (v1) at (-4,3) {$1$};
        \node[kgfignode] (v2) at (-2,3) {$2$};
        \node[kgfignode] (v3) at ( 0,3) {$3$};
        \node[kgfignode] (v4) at ( 2,3) {$4$};
        \node[kgfignode] (v5) at ( 4,3) {$5$};

        % Short edges
        \draw[line width=1.5pt] (v1) -- (v2);
        \draw[line width=1.5pt] (v2) -- (v3);
        \draw[line width=1.5pt] (v4) -- (v5);

        % Long edges, routed so that they do not overlap
        \draw[line width=1.5pt]
            (v1) .. controls (-3.0,4.3) and (1.0,4.3) .. (v4);

        \draw[line width=1.5pt]
            (v2) .. controls (-1.0,1.9) and (3.0,1.9) .. (v5);

        %------------------------------------------------
        % Skew classes
        %------------------------------------------------

        % omega_1
        \node[kgfigbox, minimum width=1.35cm, minimum height=2.5cm]
            at (-4,-0.2) {};
        \node at (-4,1.3) {$\omega_1$};
        \node at (-4,0.5) {$X_1$};
        \node at (-4,-0.2) {$Y_1$};
        \node[kgfignode, font=\normalsize] at (-4,-0.9) {$Z_1$};

        % omega_2
        \node[kgfigbox, minimum width=1.35cm, minimum height=2.5cm]
            at (-2,-0.2) {};
        \node at (-2,1.3) {$\omega_2$};
        \node at (-2,0.5) {$X_2$};
        \node at (-2,-0.2) {$Y_2$};
        \node[kgfignode, font=\normalsize] at (-2,-0.9) {$Z_2$};

        % omega_3
        \node[kgfigbox, minimum width=1.35cm, minimum height=2.5cm]
            at (0,-0.2) {};
        \node at (0,1.3) {$\omega_3$};
        \node at (0,0.5) {$X_3$};
        \node at (0,-0.2) {$Y_3$};
        \node[kgfignode, font=\normalsize] at (0,-0.9) {$Z_3$};

        % omega_4
        \node[kgfigbox, minimum width=1.35cm, minimum height=2.5cm]
            at (2,-0.2) {};
        \node at (2,1.3) {$\omega_4$};
        \node at (2,0.5) {$X_4$};
        \node at (2,-0.2) {$Y_4$};
        \node[kgfignode, font=\normalsize] at (2,-0.9) {$Z_4$};

        % omega_5
        \node[kgfigbox, minimum width=1.35cm, minimum height=2.5cm]
            at (4,-0.2) {};
        \node at (4,1.3) {$\omega_5$};
        \node at (4,0.5) {$X_5$};
        \node at (4,-0.2) {$Y_5$};
        \node[kgfignode, font=\normalsize] at (4,-0.9) {$Z_5$};

        %------------------------------------------------
        % Basis: selected elements inside the skew classes
        %------------------------------------------------
        \node at (-5.25,-0.9) {$B=$};

    \end{tikzpicture}
\end{figure}
\end{figure}
\vspace{-6mm}

The element $Z_v\in B$ is highlighted inside each skew class $\omega_v:=\lbrace X_v, Y_v, Z_v\rbrace$, where we interpret the Pauli string $B$ as a set containing all the $Z_v$ elements, see subsection~\ref{sec:notation}.
For each $v\in V$, the set $B\cup \omega_v = \lbrace Z_1, Z_2,\ldots, Z_n, X_v, Y_v\rbrace$ contains the canonical graph state generator $g_v=X_vZ_{N_v}$. See below the case of $v=2$, yielding $g_2 = ZXZIZ$ (highlighted in red).
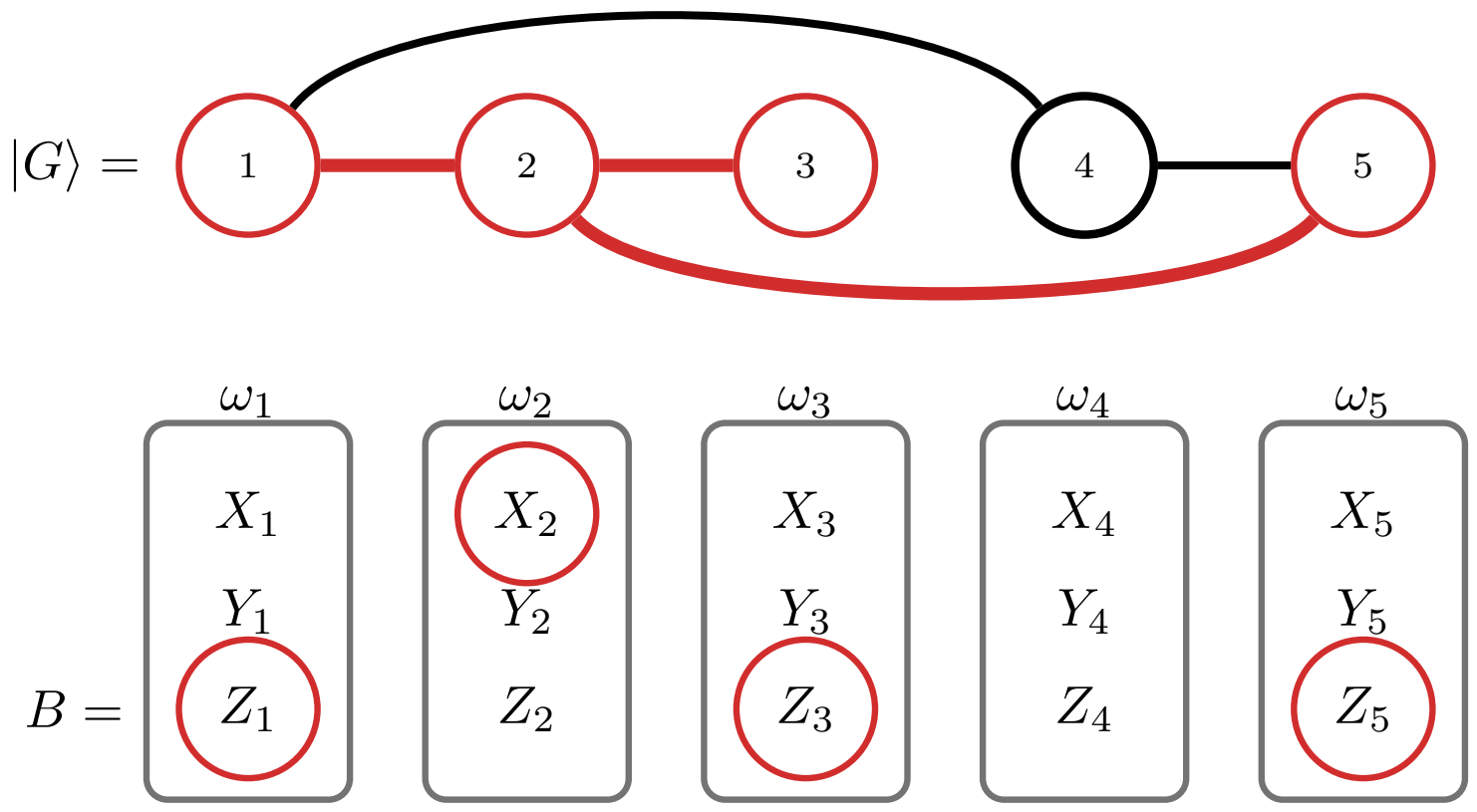
\vspace{-5mm}
\begin{figure}[H]
    \centerfloat
\begin{figure}[H]
    \centering
    \begin{tikzpicture}[scale=0.925]

        % Fundamental circuit highlight color
        \definecolor{circuitcolor}{RGB}{210,45,45}

        % Common style for highlighted circuit elements
        \tikzset{
            circuithl/.style={
                draw=circuitcolor,
                very thick,
                rounded corners=2pt,
                inner sep=3pt
            }
        }

        %------------------------------------------------
        % Graph: vertices aligned with the skew classes
        %------------------------------------------------
              \node at (-5.25,3) {$\ket{G}=$};
        \node[kgfignode,circuithl] (v1) at (-4,3) {$1$};
        \node[kgfignode,circuithl] (v2) at (-2,3) {$2$};
        \node[kgfignode,circuithl] (v3) at ( 0,3) {$3$};
        \node[kgfignode]           (v4) at ( 2,3) {$4$};
        \node[kgfignode,circuithl] (v5) at ( 4,3) {$5$};

        % Short edges
        \draw[circuitcolor, line width=2.5pt] (v1) -- (v2);
        \draw[circuitcolor, line width=2.5pt] (v2) -- (v3);
        \draw[line width=1.5pt]               (v4) -- (v5);

        % Long edges
        \draw[line width=1.5pt]
            (v1) .. controls (-3.0,4.3) and (1.0,4.3) .. (v4);

        \draw[circuitcolor, line width=2.5pt]
            (v2) .. controls (-1.0,1.9) and (3.0,1.9) .. (v5);

        %------------------------------------------------
        % Skew classes
        %------------------------------------------------

        % omega_1
        \node[kgfigbox, minimum width=1.35cm, minimum height=2.5cm]
            at (-4,-0.2) {};
        \node at (-4,1.3) {$\omega_1$};
        \node at (-4,0.5) {$X_1$};
        \node at (-4,-0.2) {$Y_1$};
        \node[kgfignode, font=\normalsize, circuithl] at (-4,-0.9) {$Z_1$};

        % omega_2
        \node[kgfigbox, minimum width=1.35cm, minimum height=2.5cm]
            at (-2,-0.2) {};
        \node at (-2,1.3) {$\omega_2$};
        \node[kgfignode, font=\normalsize, circuithl] at (-2,0.5) {$X_2$};
        \node at (-2,-0.2) {$Y_2$};
        \node at (-2,-0.9) {$Z_2$};

        % omega_3
        \node[kgfigbox, minimum width=1.35cm, minimum height=2.5cm]
            at (0,-0.2) {};
        \node at (0,1.3) {$\omega_3$};
        \node at (0,0.5) {$X_3$};
        \node at (0,-0.2) {$Y_3$};
        \node[kgfignode, font=\normalsize, circuithl] at (0,-0.9) {$Z_3$};

        % omega_4
        \node[kgfigbox, minimum width=1.35cm, minimum height=2.5cm]
            at (2,-0.2) {};
        \node at (2,1.3) {$\omega_4$};
        \node at (2,0.5) {$X_4$};
        \node at (2,-0.2) {$Y_4$};
        \node at (2,-0.9) {$Z_4$};

        % omega_5
        \node[kgfigbox, minimum width=1.35cm, minimum height=2.5cm]
            at (4,-0.2) {};
        \node at (4,1.3) {$\omega_5$};
        \node at (4,0.5) {$X_5$};
        \node at (4,-0.2) {$Y_5$};
        \node[kgfignode, font=\normalsize, circuithl] at (4,-0.9) {$Z_5$};

        %------------------------------------------------
        % Basis
        %------------------------------------------------
        \node at (-5.25,-0.9) {$B=$};

    \end{tikzpicture}
\end{figure}
\end{figure}
\vspace{-6mm}
 In fact, $g_v$ is the \emph{unique} non-identity stabilizer in $ B\cup \omega_v$. Note that the uniqueness of $g_v$ implies that the only stabilizers contained in $g_v$ itself are the identity and $g_v$ itself. Stabilizers such that the only stabilizers contained in them are themselves and the identity are called \emph{circuits}\footnote{This terminology is taken from matroid theory, explaining the unfortunate clash with the notion of a quantum circuit.}, and the reader might find it helpful to compare this definition with that of a prime number\footnote{Or an \emph{irreducible element} if you are a number theorist.}. In particular, we will call $g_v$ the \emph{fundamental circuit} with respect to $B$ and $v$. Note that the graph $G$ can be reconstructed from the fundamental circuits $\lbrace g_v \rbrace_{v\in V}$, since the support of each $g_v$ is just the closed neighborhood $\overline{N}_v$ of $v$.

\subsection{Bases}
In the previous example we showed how one can recover the canonical graph state generator $g_v=X_vZ_{N_v}$ as the unique circuit contained in $B\cup \omega_v$, when $B=Z^{\otimes n}$.
So far, this appears to be a rather contrived way to think of the canonical generators of a graph state. This approach has the benefit, however, that it extends to arbitrary stabilizer states $\ket{\psi}$, and more general Pauli strings than $Z^{\otimes n}$. The Pauli strings we consider are called \emph{(matroid) bases} of the state $\ket{\psi}$ and are defined in the following manner.

\begin{definition}[Bases of stabilizer states]\label{def:bases}
Let $\ket{\psi}$ be a stabilizer state. A Pauli string $B$ is a basis of $\ket{\psi}$ if $B$ is full-weight, and no non-empty subset of $B$ is a stabilizer. We denote the set of all bases of $\ket{\psi}$ by $\mathcal{B}_\psi$.
\end{definition}

When the state under consideration is unimportant or clear from the context, we will write $\mathcal{B}=\mathcal{B}_\psi$. Here are some examples of bases.

\begin{example}\label{example:basis1}
Let $\ket{\psi_1}=\ket{00}$. The stabilizer group of $\ket{\psi_1}$ is, up to phases, given by $\lbrace{II, IZ, ZI, ZZ\rbrace}$. Then $YX$ is a basis, since none of $YX, IX, YI$ is a stabilizer. Extending this idea, we find that $\mathcal{B}_{\psi_1} = \lbrace{XX, XY, YX, YY\rbrace}$ 
since these are all the full-weight Pauli strings containing neither $Z_1$ nor $Z_2$.
\end{example}
\begin{example}\label{example:basis12}
Let $\ket{\psi_2}\propto\ket{00}+\ket{11}$, whose phaseless stabilizer group consists of all Pauli strings whose first and second entries agree, i.e.~$\lbrace II, XX, YY, ZZ\rbrace$. We find that $ \mathcal{B}_{\psi_2} = \lbrace{XZ, XY, YX, YZ, ZX, ZY\rbrace}$, since these are precisely the full-weight Pauli strings whose two entries disagree.
\end{example}

\begin{example}\label{example:basis2}
Let $\ket{G}$ be any $n$-qubit graph state. Every non-identity stabilizer of $\ket{G}$ is a non-empty product of its canonical generators $X_vZ_{N_v}$. But any such product needs to contain at least one $X$ or $Y$, so $Z^{\otimes n}$ has no non-identity stabilizer as a subset. That is, $Z^{\otimes n} \in \mathcal{B}_{G}$, independent of the graph $G$.  
\end{example}

\begin{example}
A slightly tedious calculation shows that the bases of the complete graph state $\ket{K_3}$ on three vertices (which is locally equivalent to a $3$-GHZ state) are given by 

\[
\mathcal{B}_{K_3}=\left\lbrace
\begin{array}{cccc}
XZX& YZX& XZY&  XYX\nonumber\\
ZXX& ZYX& ZXY&  YXX\nonumber\\
XXZ& XYZ& YXZ&  XXY\nonumber\\
ZZZ& ZZY& ZYZ&  YZZ\nonumber 
\end{array}\right\rbrace .
\]
\end{example}

Bases can be interpreted as Pauli measurements that yield maximally random outcomes. To make this concrete, assume the setting where each qubit of a stabilizer state is separated in space, such that only local measurements can be performed. A Pauli string $P$ then specifies a measurement in the following manner: perform the local Pauli measurement $P_v$ for each qubit $v\in \textrm{supp}(P)$, where $\textrm{supp}(P)$ denotes the support of $P$.
Bases then have the following clean physical interpretation: measuring each qubit $v$ according to $B_v$ will yield all $2^n$ possible measurement outcomes uniformly at random, see~\cite{goodenough2026asymptotic}. In particular, this means there are no correlations whatsoever among the outcomes. 

\begin{property}[\cite{goodenough2026asymptotic}]\label{prop:bases}
The bases $B\in \mathcal{B}_\psi$ correspond precisely to the Pauli measurements on all qubits that yield completely random outcomes.
\end{property}

\begin{example}
Measuring $\ket{00}$ in $P$ will yield completely random outcomes iff no $Z$-measurement is performed. Two parties measuring $\ket{00}+\ket{11}$ will see correlations iff they measure the same Pauli operator. This reconstructs the bases from examples \ref{example:basis1} and \ref{example:basis12}.
\end{example}

\newpage
\begin{example}
Measuring every qubit of a graph state in the computational basis yields completely random outcomes. This can be seen from the well-known expansion of graph states in the computational basis~\cite{hoyer2006resources}:
\begin{align}
\ket{G}= \frac{1}{\sqrt{2^n}}\sum_{b\in \lbrace{0,1\rbrace}^n}\left(-1\right)^{\frac{b^TAb}{2}}\ket{b},
\end{align}
where $A$ is the adjacency matrix of $G$.
Since this is a superposition over all computational basis vectors whose amplitudes all have equal magnitude $\frac{1}{\sqrt{2^n}}$, measuring every qubit in $Z$ yields a uniformly random computational basis state. This reconstructs the statement from example \ref{example:basis2}, namely that $Z^{\otimes n}$ is a basis for every graph state $\ket{G}$.
\end{example}

The term `basis' comes from matroid theory, where they generalize the notion of a basis in linear algebra. Note that a basis in linear algebra is a set of vectors $B$ that is linearly independent, and for which it is impossible to add any other vector to $B$ such that the new set remains linearly independent. Note how this is analogous to bases of stabilizer states: a basis $B$ is a collection of measurements such that 1) \emph{the measurement outcomes are independent}, in the sense that all possible bitstrings are equally likely as outcome, and 2) for which one cannot add any other measurements, for the simple fact that all qubits have been measured already. For more information on this analogy, see~\cite{goodenough2026asymptotic, bouchet1997multimatroids}.

At first sight, bases appear to be rather useless measurements, since they yield completely random outcomes. In the next subsection, however, we show that they provide a natural abstraction of graphs up to local complementations.

\subsection{Fundamental graphs of bases}\label{sec:fund_graphs_of_bases}
Generalizing the example from subsection \ref{sec:a_simple}, we now show how to assign a \emph{fundamental graph} $G_B$ to each basis $B\in \mathcal{B}_\psi$.

We will deal with the general case, but show as a running example in the figures the case of a three-qubit GHZ state $\ket{\textrm{GHZ}}\propto \ket{000}+\ket{111}$ and the basis $B=XYZ$. This state has, up to phases, the following stabilizer group

\[\begin{array}{cc}
III & XXX\\
IZZ & XYY\\
ZIZ & YXY\\
ZZI & YYX\\
\end{array}\, ,\]
from which it can be checked that $B=XYZ$ is indeed a basis.

To start, let $\ket{\psi}$ be an arbitrary stabilizer state, and $B\in \mathcal{B}_\psi$ an arbitrary basis.
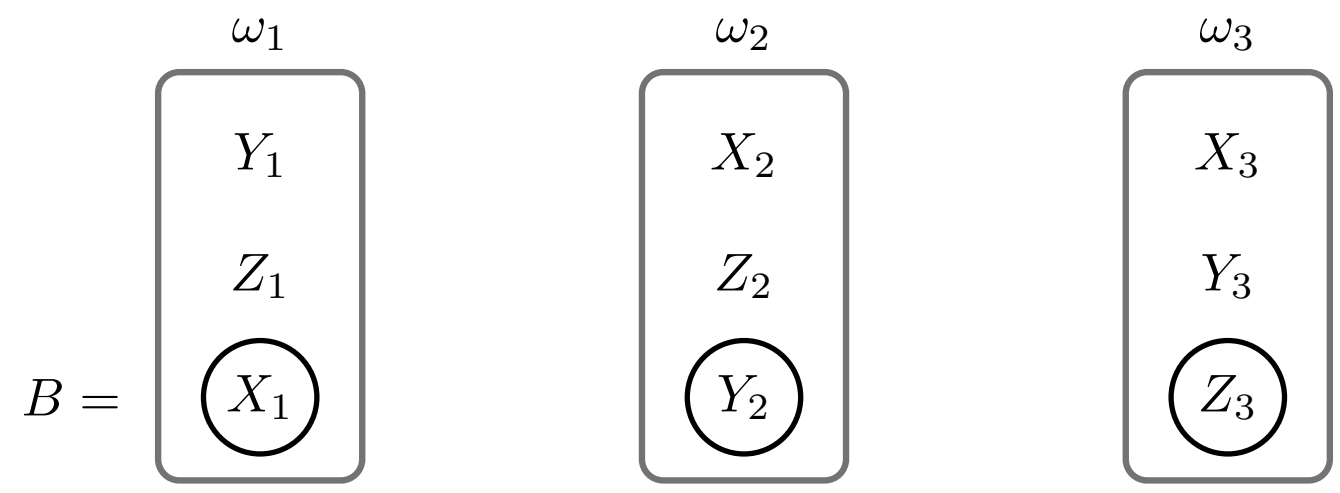
\vspace{-3mm}
\begin{figure}[H]
    \centerfloat
\begin{figure}[H]
    \centering
    \begin{tikzpicture}

        %------------------------------------------------
        % Controllable vertical spacing between elements
        %------------------------------------------------
        \def\nodesep{0.80}

        %------------------------------------------------
        % Circle style for basis elements
        %------------------------------------------------
        \tikzset{
            bsel/.style={
                draw,
                circle,
                line width=1pt,
                minimum size=7.5mm,
                inner sep=0pt
            }
        }

        %------------------------------------------------
        % Skew class omega_1
        % Basis element X_1 is kept at the bottom.
        %------------------------------------------------
        \node[kgfigbox, minimum width=1.35cm, minimum height=2.7cm]
            at (-3.2,-0.2) {};
        \node at (-3.2,1.4) {$\omega_1$};

        \node[font=\normalsize] at (-3.2,{-0.2+\nodesep}) {$Y_1$};

        \node[font=\normalsize] at (-3.2,-0.2) {$Z_1$};

        \node[font=\normalsize] at (-3.2,{-0.2-\nodesep}) {$X_1$};
        \node[bsel] at (-3.2,{-0.2-\nodesep}) {};

        %------------------------------------------------
        % Skew class omega_2
        % Basis element Y_2 is kept at the bottom.
        %------------------------------------------------
        \node[kgfigbox, minimum width=1.35cm, minimum height=2.7cm]
            at (0,-0.2) {};
        \node at (0,1.4) {$\omega_2$};

        \node[font=\normalsize] at (0,{-0.2+\nodesep}) {$X_2$};

        \node[font=\normalsize] at (0,-0.2) {$Z_2$};

        \node[font=\normalsize] at (0,{-0.2-\nodesep}) {$Y_2$};
        \node[bsel] at (0,{-0.2-\nodesep}) {};

        %------------------------------------------------
        % Skew class omega_3
        % Basis element Z_3 is kept at the bottom.
        %------------------------------------------------
        \node[kgfigbox, minimum width=1.35cm, minimum height=2.7cm]
            at (3.2,-0.2) {};
        \node at (3.2,1.4) {$\omega_3$};

        \node[font=\normalsize] at (3.2,{-0.2+\nodesep}) {$X_3$};

        \node[font=\normalsize] at (3.2,-0.2) {$Y_3$};

        \node[font=\normalsize] at (3.2,{-0.2-\nodesep}) {$Z_3$};
        \node[bsel] at (3.2,{-0.2-\nodesep}) {};

        %------------------------------------------------
        % Basis
        %------------------------------------------------
        \node at (-4.45,{-0.2-\nodesep}) {$B=$};

        %------------------------------------------------
        % Generator labels
        %------------------------------------------------
        % \node at (-3.2,-2.4)
        %     {$g_1^{XYZ}=ZIZ$};

        % \node at (0,-2.4)
        %     {$g_2^{XYZ}=IZZ$};

        % \node at (3.2,-2.4)
        %     {$g_3^{XYZ}=XYY$};

    \end{tikzpicture}
\end{figure}
\end{figure}
\vspace{-3mm}
For each qubit $v$, the fundamental circuit $g_v^B$ of $\ket{\psi}$ with respect to $B$ and $v$ is the unique circuit contained in $B\cup \omega_v$. This is the exact same definition as from section~\ref{sec:fundamental_graphs}, but extended to arbitrary bases and stabilizer states.

\newpage
\begin{definition}[Fundamental circuits with respect to a basis $B$]\label{def:fund_circuits}
Let $B$ be a basis of a stabilizer state $\ket{\psi}$. The \emph{fundamental circuit} $g_v^B$ (with respect to $B$, $v$ and $\ket{\psi}$) is the unique non-identity stabilizer contained in $B\cup \omega_v$. The existence and uniqueness of $g_v^B$ follow from theorem 4.2 from~\cite{bouchet2001multimatroids}.
\end{definition}

We show below the fundamental circuits of $\ket{\textrm{GHZ}}$ with respect to $B$.
% \vspace{-3mm}
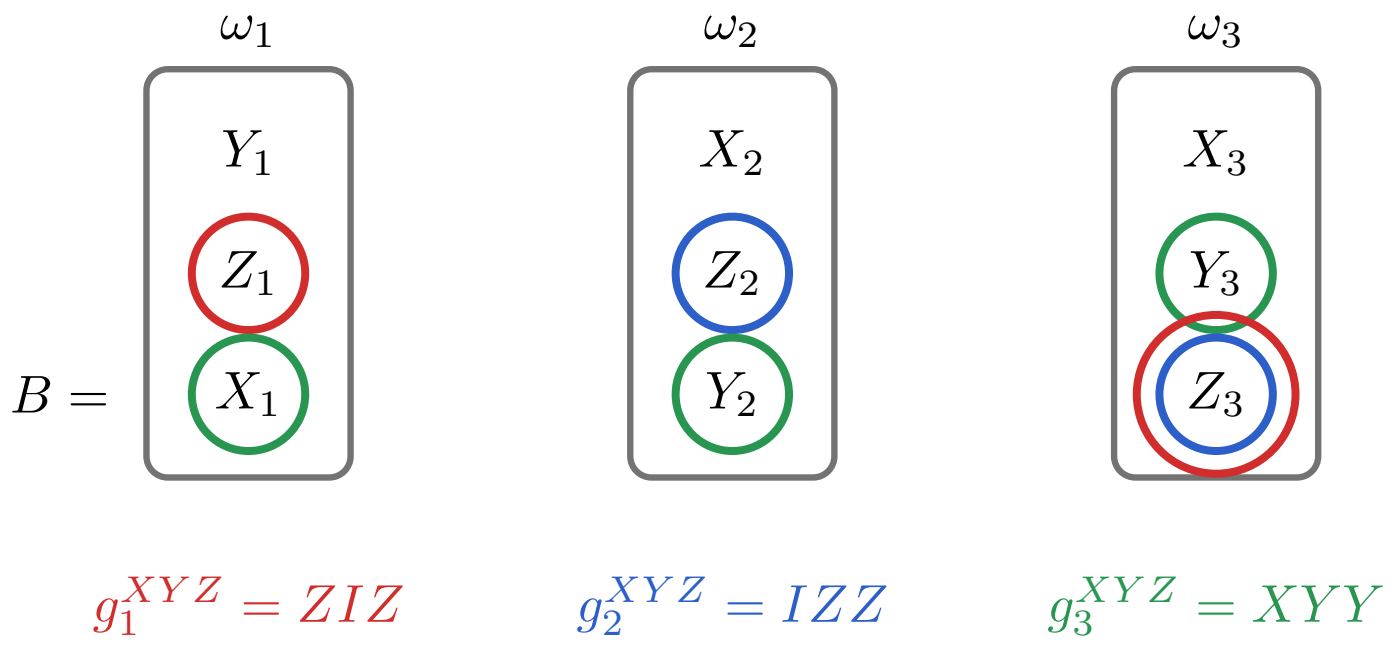
\begin{figure}[H]
    \centerfloat
\begin{figure}[H]
    \centering
    \begin{tikzpicture}

        %------------------------------------------------
        % Controllable vertical spacing between elements
        %------------------------------------------------
        \def\nodesep{0.80}

        %------------------------------------------------
        % Fundamental circuit colors
        %------------------------------------------------
        \definecolor{circuitone}{RGB}{210,45,45}
        \definecolor{circuittwo}{RGB}{45,95,200}
        \definecolor{circuitthree}{RGB}{40,150,80}

        % Circular highlights
        \tikzset{
            c1/.style={
                draw=circuitone,
                circle,
                line width=1.5pt,
                minimum size=7.5mm,
                inner sep=0pt
            },
            c2/.style={
                draw=circuittwo,
                circle,
                line width=1.5pt,
                minimum size=7.5mm,
                inner sep=0pt
            },
            c3/.style={
                draw=circuitthree,
                circle,
                line width=1.5pt,
                minimum size=7.5mm,
                inner sep=0pt
            },
            c1big/.style={
                draw=circuitone,
                circle,
                line width=1.5pt,
                minimum size=10.5mm,
                inner sep=0pt
            }
        }

        %------------------------------------------------
        % Skew class omega_1
        % Basis element X_1 is kept at the bottom.
        %------------------------------------------------
        \node[kgfigbox, minimum width=1.35cm, minimum height=2.7cm]
            at (-3.2,-0.2) {};
        \node at (-3.2,1.4) {$\omega_1$};

        \node[font=\normalsize] at (-3.2,{-0.2+\nodesep}) {$Y_1$};

        \node[font=\normalsize] at (-3.2,-0.2) {$Z_1$};
        \node[c1] at (-3.2,-0.2) {};

        \node[font=\normalsize] at (-3.2,{-0.2-\nodesep}) {$X_1$};
        \node[c3] at (-3.2,{-0.2-\nodesep}) {};

        %------------------------------------------------
        % Skew class omega_2
        % Basis element Y_2 is kept at the bottom.
        %------------------------------------------------
        \node[kgfigbox, minimum width=1.35cm, minimum height=2.7cm]
            at (0,-0.2) {};
        \node at (0,1.4) {$\omega_2$};

        \node[font=\normalsize] at (0,{-0.2+\nodesep}) {$X_2$};

        \node[font=\normalsize] at (0,-0.2) {$Z_2$};
        \node[c2] at (0,-0.2) {};

        \node[font=\normalsize] at (0,{-0.2-\nodesep}) {$Y_2$};
        \node[c3] at (0,{-0.2-\nodesep}) {};

        %------------------------------------------------
        % Skew class omega_3
        % Basis element Z_3 is kept at the bottom.
        %------------------------------------------------
        \node[kgfigbox, minimum width=1.35cm, minimum height=2.7cm]
            at (3.2,-0.2) {};
        \node at (3.2,1.4) {$\omega_3$};

        \node[font=\normalsize] at (3.2,{-0.2+\nodesep}) {$X_3$};

        \node[font=\normalsize] at (3.2,-0.2) {$Y_3$};
        \node[c3] at (3.2,-0.2) {};

        \node[font=\normalsize] at (3.2,{-0.2-\nodesep}) {$Z_3$};

        % Z_3 belongs to both C_1 and C_2
        \node[c2]    at (3.2,{-0.2-\nodesep}) {};
        \node[c1big] at (3.2,{-0.2-\nodesep}) {};

        %------------------------------------------------
        % Basis
        %------------------------------------------------
        \node at (-4.45,{-0.2-\nodesep}) {$B=$};

        %------------------------------------------------
        % Circuit labels
        %------------------------------------------------
        \node[text=circuitone] at (-3.2,-2.4)
            {$g_1^{XYZ}=ZIZ$};

        \node[text=circuittwo] at (0,-2.4)
            {$g_2^{XYZ}=IZZ$};

        \node[text=circuitthree] at (3.2,-2.4)
            {$g_3^{XYZ}=XYY$};

    \end{tikzpicture}
\end{figure}
\vspace*{-2mm}
\end{figure}
With the fundamental circuits with respect to $B$ in hand, we can define the fundamental graph $G_B$.

\begin{definition}[Fundamental graph of a basis $B$~\cite{bouchet2001multimatroids}]\label{def:fundamental_graph}
    The fundamental graph $G_B$ of $\ket{\psi}$ with respect to $B$ is the graph with vertex set $V$, where each $v$ has as
    its closed neighborhood $\overline{N}_v:=N_v\cup \lbrace v\rbrace $ equal to $\textrm{supp}\left(g_v^B\right)$.
    Equivalently, two distinct qubits $v, w\in V$ are adjacent $\iff$ $B_w \in g_v^B$. This equivalence follows from the fact that $w\in \textrm{supp}(g_v^B)\setminus \lbrace v \rbrace \iff  \left(g_v^B\right)_w = B_w$. 
\end{definition}

% The fundamental graph $G_B$ has a concrete physical interpration: when measuring according to $B$, random outcomes are obtained. There are two changing the measurement at $v$

For the example of a GHZ state $\ket{000}+\ket{111}$ and $B=XYZ$, $G_B$ is the star graph with vertex $3$ as its center. This follows from the fact that $\textrm{supp}(g_1^{XYZ})=\lbrace 1, 3\rbrace$, $\textrm{supp}(g_2^{XYZ})=\lbrace 2, 3\rbrace$, and $\textrm{supp}(g_3^{XYZ})=\lbrace 1, 2, 3\rbrace$.

The choice of basis $B$ in the above was arbitrary. What happens if we choose a different basis $B'$? As an example, the fundamental graph $G_{B'}$ with $B'=XYX$  for $\ket{\textrm{GHZ}}$ is the complete graph on $3$ vertices. This follows since, with respect to this new basis $B'$, the fundamental circuits all have full support:
% \vspace{-3.5mm}
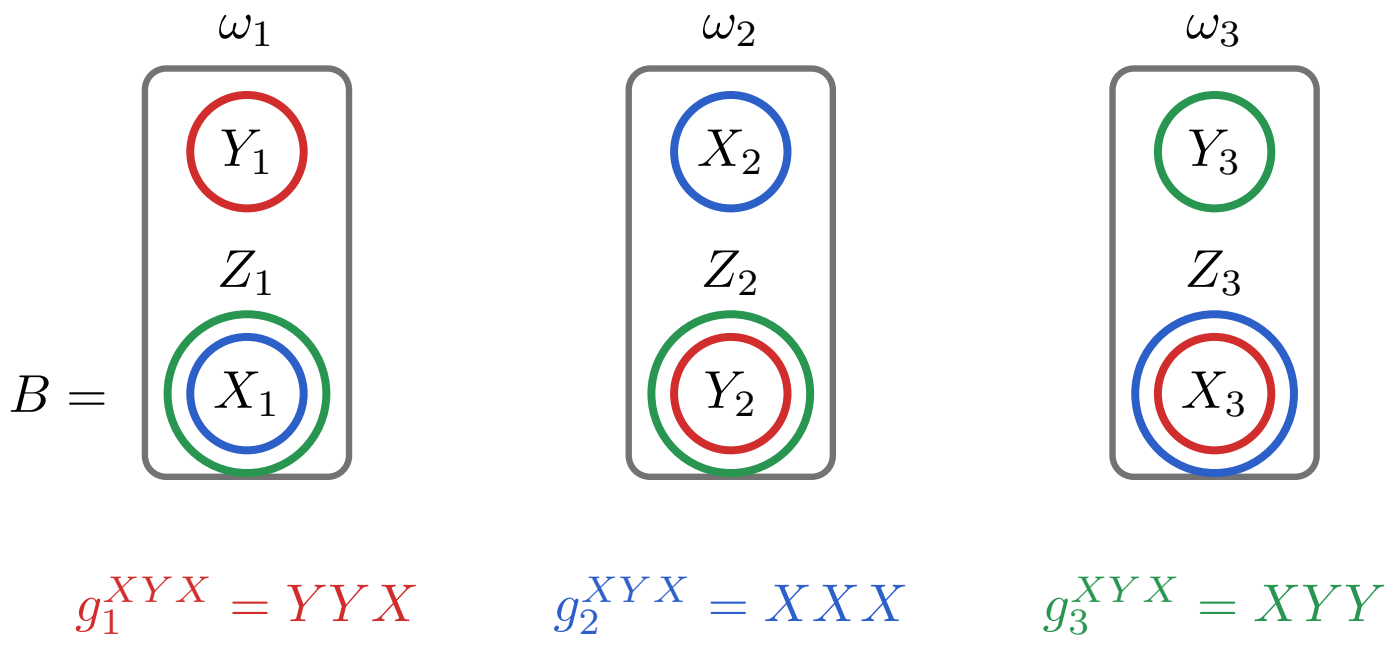
\begin{figure}[H]
    \centerfloat
\begin{figure}[H]
    \centering
    \begin{tikzpicture}

        %------------------------------------------------
        % Controllable vertical spacing between elements
        %------------------------------------------------
        \def\nodesep{0.80}

        %------------------------------------------------
        % Fundamental circuit colors
        %------------------------------------------------
        \definecolor{circuitone}{RGB}{210,45,45}
        \definecolor{circuittwo}{RGB}{45,95,200}
        \definecolor{circuitthree}{RGB}{40,150,80}

        % Circular highlights
        \tikzset{
            c1/.style={
                draw=circuitone,
                circle,
                line width=1.5pt,
                minimum size=7.5mm,
                inner sep=0pt
            },
            c2/.style={
                draw=circuittwo,
                circle,
                line width=1.5pt,
                minimum size=7.5mm,
                inner sep=0pt
            },
            c3/.style={
                draw=circuitthree,
                circle,
                line width=1.5pt,
                minimum size=7.5mm,
                inner sep=0pt
            },
            c1big/.style={
                draw=circuitone,
                circle,
                line width=1.5pt,
                minimum size=10.5mm,
                inner sep=0pt
            },
            c2big/.style={
                draw=circuittwo,
                circle,
                line width=1.5pt,
                minimum size=10.5mm,
                inner sep=0pt
            },
            c3big/.style={
                draw=circuitthree,
                circle,
                line width=1.5pt,
                minimum size=10.5mm,
                inner sep=0pt
            }
        }

        %------------------------------------------------
        % Skew class omega_1
        %
        % Basis element X_1 is kept at the bottom.
        % Remaining elements are ordered as Y, Z.
        %------------------------------------------------
        \node[kgfigbox, minimum width=1.35cm, minimum height=2.7cm]
            at (-3.2,-0.2) {};
        \node at (-3.2,1.4) {$\omega_1$};

        \node[font=\normalsize] at (-3.2,{-0.2+\nodesep}) {$Y_1$};
        \node[c1] at (-3.2,{-0.2+\nodesep}) {};

        \node[font=\normalsize] at (-3.2,-0.2) {$Z_1$};

        \node[font=\normalsize] at (-3.2,{-0.2-\nodesep}) {$X_1$};
        % X_1 belongs to C_2 and C_3
        \node[c2]    at (-3.2,{-0.2-\nodesep}) {};
        \node[c3big] at (-3.2,{-0.2-\nodesep}) {};

        %------------------------------------------------
        % Skew class omega_2
        %
        % Basis element Y_2 is kept at the bottom.
        % Remaining elements are ordered as X, Z.
        %------------------------------------------------
        \node[kgfigbox, minimum width=1.35cm, minimum height=2.7cm]
            at (0,-0.2) {};
        \node at (0,1.4) {$\omega_2$};

        \node[font=\normalsize] at (0,{-0.2+\nodesep}) {$X_2$};
        \node[c2] at (0,{-0.2+\nodesep}) {};

        \node[font=\normalsize] at (0,-0.2) {$Z_2$};

        \node[font=\normalsize] at (0,{-0.2-\nodesep}) {$Y_2$};
        % Y_2 belongs to C_1 and C_3
        \node[c1]    at (0,{-0.2-\nodesep}) {};
        \node[c3big] at (0,{-0.2-\nodesep}) {};

        %------------------------------------------------
        % Skew class omega_3
        %
        % Basis element X_3 is kept at the bottom.
        % Remaining elements are ordered as Y, Z.
        %------------------------------------------------
        \node[kgfigbox, minimum width=1.35cm, minimum height=2.7cm]
            at (3.2,-0.2) {};
        \node at (3.2,1.4) {$\omega_3$};

        \node[font=\normalsize] at (3.2,{-0.2+\nodesep}) {$Y_3$};
        \node[c3] at (3.2,{-0.2+\nodesep}) {};

        \node[font=\normalsize] at (3.2,-0.2) {$Z_3$};

        \node[font=\normalsize] at (3.2,{-0.2-\nodesep}) {$X_3$};
        % X_3 belongs to C_1 and C_2
        \node[c1]    at (3.2,{-0.2-\nodesep}) {};
        \node[c2big] at (3.2,{-0.2-\nodesep}) {};

        %------------------------------------------------
        % Basis
        %------------------------------------------------
        \node at (-4.45,{-0.2-\nodesep}) {$B=$};

        %------------------------------------------------
        % Circuit labels
        %------------------------------------------------
        \node[text=circuitone] at (-3.2,-2.4)
            {$g_1^{XYX}=YYX$};

        \node[text=circuittwo] at (0,-2.4)
            {$g_2^{XYX}=XXX$};

        \node[text=circuitthree] at (3.2,-2.4)
            {$g_3^{XYX}=XYY$};

    \end{tikzpicture}
\end{figure}
\vspace*{-2mm}
\end{figure}
\vspace{-3mm}
 Comparing $G_B$ and $G_{B'}$, we find them to be equivalent after a local complementation at vertex $3$. This is not a coincidence, as the following theorem shows.

\newpage
\begin{theorem}[\cite{bouchet1988graphic}]\label{theorem:bases_surj_into_graphs}
Let $\ket{\psi}$ be a stabilizer state and $H$ a graph. Then
\[\ket{\psi} \simeq_{\textrm{LC}}\ket{H} \iff H=G_B\textrm{ is a fundamental graph for some basis } B\in\mathcal{B}_\psi .\]
\end{theorem}
\begin{proof}
We first prove the forward direction, i.e.~that $\ket{\psi} \simeq_{\textrm{LC}}\ket{H}$ implies that $H$ is a fundamental graph for some basis $B\in\mathcal{B}_\psi$. Let $C$ be a local Clifford mapping $\ket{\psi}$ to $\ket{H}$. Since $Z^{\otimes n}$ is a basis for $\ket{H}$ (see example \ref{example:basis2}), $B:=C^{\dagger}\left(Z^{\otimes n}\right)C
$ is a basis for $\ket{\psi}$.
\medskip

We claim that $G_B=H$. To see this, first note that the definition of a fundamental graph depends only on the support of its fundamental circuits. In turn, the support of a fundamental circuit does not change when simultaneously applying local Cliffords to both the state and the basis. This follows from the fact that local Cliffords only locally relabel Paulis, and that the fundamental circuits $g_v^B$ are defined as being the unique circuits in $B\cup \omega_v$. Finally, we use that the fundamental graph of a graph state $\ket{H}$ with respect to the basis $Z^{\otimes n}$ equals $H$. This proves the forward direction.

For the other direction, since the fundamental circuits $g_v^B$ generate the whole stabilizer group, it suffices to show that all $g_v^B$ can be simultaneously transformed to the form of a graph state canonical generator, i.e.~$X_{v}Z_{N_v}$, where the neighborhoods $N_v$ are with respect to the fundamental graph $G_B$. A simple verification shows that the following (unique) local Clifford works: for each qubit $v$, take $B_v\mapsto Z_v$ and $\left(g_v^B\right)_v\mapsto X_v$. This map is well-defined since $\left(g_v^B\right)_v$ and $B_v$ are distinct. This follows from the fact that $g_v^B$ is the unique circuit in $B\cup \omega_v$ and $B$ itself does not contain $g_v^{B}$ since it is a basis (see definition \ref{def:bases}).
\end{proof}

This theorem states that the fundamental graphs associated to a state $\ket{\psi}$ are precisely the graphs whose graph states are locally equivalent to $\ket{\psi}$. Since a fundamental graph $G_B$ is defined relative to a basis $B\in \mathcal{B}$, it is natural to wonder if local complementations have an analogue in terms of bases. As a hint for what is to come, observe that for the example of the GHZ state, the two bases $B=XYZ$ and $B'=XYX$ differ only in their third entry, while the two fundamental graphs $G_B$ and $G_{B'}$ are related by a local complementation on the third vertex.

\newpage
\section{Local complementations as switching on bases}\label{sec:local_comps_as_switches}
Recall that we are looking to complete the following diagram,

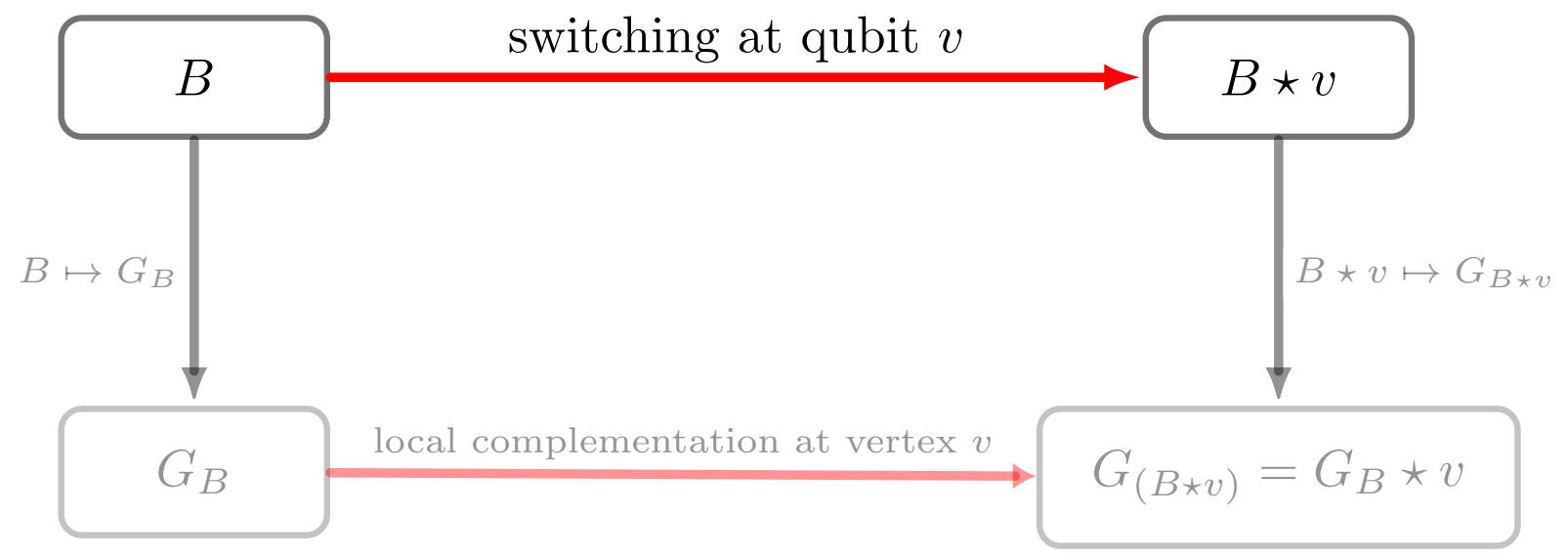
\begin{figure}[H]
    \centerfloat
% Horizontal separation between left- and right-hand columns
\newcommand{\kghorizsep}{55mm}

% Opacity of everything except the top row and switching arrow
% 1 = fully opaque, 0 = invisible
\newcommand{\kgloweropacity}{0.425}

\begin{figure}[H]
    \centering
    \begin{tikzpicture}[node distance=18mm and \kghorizsep]

        %--------------------------------------------------
        % Fully opaque top row
        %--------------------------------------------------
        \node[kgfigbox, minimum width=18mm] (B) {$B$};
        \node[kgfigbox, minimum width=18mm, right=of B] (Bs) {$B\star v$};

        \draw[kgfigqone,-{Latex[length=2.6mm,width=1.8mm]}]
            (B) -- node[
                above,
                inner sep=2pt,
                font=\normalsize
            ]{switching at qubit $v$} (Bs);

        %--------------------------------------------------
        % Everything else has tunable opacity
        %--------------------------------------------------
        \begin{scope}[opacity=\kgloweropacity]

            \node[kgfigbox, minimum width=18mm, below=of B]
                (GB) {$G_B$};
            \node[kgfigbox, minimum width=25mm, below=of Bs]
                (GBs) {$G_{(B\star v)}=G_{B}\star v$};

            \draw[kgfigarrow]
                (B) -- node[
                    left,
                    fill=white,
                    inner sep=2pt,
                    font=\scriptsize
                ]{$B\mapsto G_B$} (GB);

            \draw[kgfigarrow]
                (Bs) -- node[
                    right,
                    fill=white,
                    inner sep=2pt,
                    font=\scriptsize
                ]{$B\star v\mapsto G_{B\star v}$} (GBs);

            \draw[kgfigqone,-{Latex[length=1.6mm,width=1.8mm]}]
                (GB) -- node[
                    above,
                    inner sep=2pt,
                    font=\scriptsize
                ]{local complementation at vertex $v$} (GBs);

        \end{scope}

    \end{tikzpicture}
\end{figure}
\end{figure}

Constructing the fundamental graph $G_B$ from a given basis $B$ was discussed in the previous section, so it remains to define the appropriate operation $B\mapsto B\star v$ on bases in the above diagram. We refer to this operation as \emph{switching} $B$ at $v$~\cite{bouchet2001multimatroids}.

Two natural constraints for a switching $B\mapsto B\star v$ are to 1) demand that $B\star v$ is a basis, and 2) that $B$ and $B\star v$ differ only in the entry at $v$. The following theorem by Bouchet---which we reformulate and reprove in quantum information terms---states that these constraints are enough to \emph{define} $B\star v$.

\begin{definition}
Two full-weight Pauli strings $P, P'$ are said to be neighboring at $v$ if for all $i\neq v$, $P_i=P_i'$, and $P_v\neq P_v'$.
\end{definition}

\begin{theorem}[Uniqueness of bases neighboring at a qubit $v$]\label{theorem:switching}
Let $B\in \mathcal{B}$ be a basis of a stabilizer state $\ket{\psi}$, and let $v$ be an arbitrary qubit of $\ket{\psi}$. Then there exists a unique basis $B'$ in $\mathcal{B}$ such that $B$ and $B'$ are neighboring at $v$.
\end{theorem}

We will prove this shortly. In the meantime, theorem~\ref{theorem:switching} allows us to define switchings on bases.

\begin{definition}[Switching on bases]\label{def:switch_basis}
Let $B\in \mathcal{B}$ be a basis of a stabilizer state $\ket{\psi}$, and let $v$ be an arbitrary qubit of $\ket{\psi}$. The \emph{switching} of $B$ at $v$ is the unique basis $B'\in \mathcal{B}$ that neighbors $B$ at $v$. We denote $B'$ by $B\star v$, overloading the same notation as used for local complementations.
\end{definition}

\begin{example}
Recall from examples~\ref{example:basis1} and \ref{example:basis12} that the bases of $\ket{00}$ and $\ket{00}+\ket{11}$ are given by $\mathcal{B}_{\psi_1}=\lbrace{XX, XY, YX, YY\rbrace}$ and $ \mathcal{B}_{\psi_2} = \lbrace{XZ, XY, YX, YZ, ZX, ZY\rbrace}$, respectively. For $\ket{00}$, switching $YX$ at the first qubit yields $XX$. For $\ket{00}+\ket{11}$, switching $YX$ at the first qubit yields $ZX$, while switching at the second qubit yields $YZ$.
\end{example}

Before proving theorem \ref{theorem:switching}, we will need the following incredibly useful property of stabilizer states under Pauli measurements~\cite{gottesman1997stabilizer, aaronson2004improved}. We provide a proof in the appendix for completeness, see~\ref{property:stab_no_signs_meas2}.

\begin{property}[Stabilizer states are well-behaved under Pauli measurements]\label{property:stab_no_signs_meas}
Let $\ket{\psi}$ be any stabilizer state, and perform any fixed sequence of Pauli measurements on $\ket{\psi}$. Let $\ket{\phi, O}$ and $\ket{\phi, O'}$ be any two post-measurement states after this process, where $O$ and $O'$ denote two possible sets of measurement outcomes. Then $\ket{\phi, O}$ and $\ket{\phi, O'}$ are equivalent up to Pauli unitaries; in particular, $\ket{\phi, O}$ is an eigenstate of a Pauli operator $Q$ $\iff$ $\ket{\phi, O'}$ is an eigenstate of $Q$.
\end{property}

As a simple example, measuring the first qubit of $\ket{00}+\ket{11}$ in $Z$ will collapse the second qubit to either $\ket{0}$ or $\ket{1}$ depending on the measurement outcome. Property~\ref{property:stab_no_signs_meas} now allows for a simple proof of theorem \ref{theorem:switching}.

\begin{proof}[Proof of theorem \ref{theorem:switching}]
We want to prove that, for every basis $B\in \mathcal{B}$ and every qubit $v$, there exists a unique basis $B'\in \mathcal{B}$ such that $B$ and $B'$ are neighboring at $v$.

Let $\ket{\psi}$ be an arbitrary stabilizer state, with qubits $1, 2, \ldots, v, \ldots, n$. We visualize this as follows.

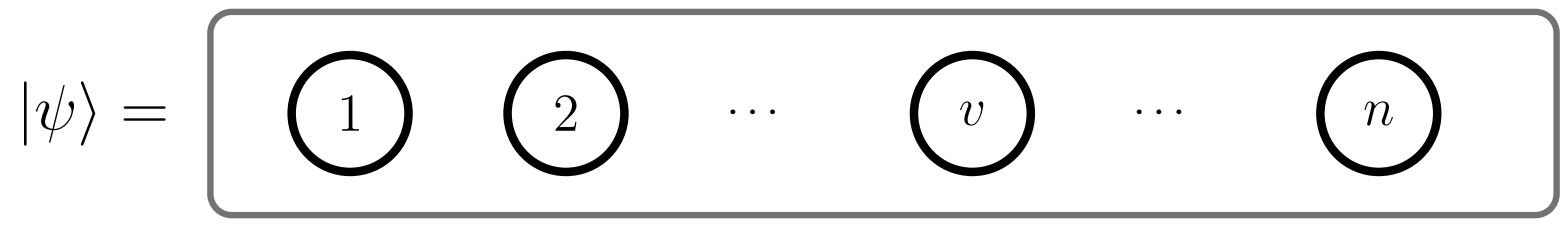
\begin{figure}[H]
    \centerfloat
\begin{figure}[H]
    \centering
    \begin{tikzpicture}[scale=0.84,transform shape]

        %------------------------------------------------
        % Adjustable measurement parameters
        %------------------------------------------------
        \def\meassize{10mm}        % size of measurement box
        \def\measarcradius{2.8mm}  % radius of meter arc
        \def\measneedle{2.5mm}     % length/height of meter needle

        %------------------------------------------------
        % Styles
        %------------------------------------------------
        \tikzset{
            measbox/.style={
                draw,
                minimum width=\meassize,
                minimum height=\meassize,
                inner sep=0pt
            },
            measlabel/.style={
                font=\normalsize
            },
            outcomelabel/.style={
                font=\normalsize
            }
        }

        %------------------------------------------------
        % State box containing the qubits
        %------------------------------------------------
        \node[
            kgfigbox,
            minimum width=10.6cm,
            minimum height=1.6cm
        ] (psi) at (0,1.35) {};

        % State label on the left
        \node[
            anchor=east,
            font=\Large
        ] at (-5.5,1.35) {$\ket{\psi}=$};

        % Invisible matching label on the right,
        % so that the full figure remains centered
        \node[
            anchor=west,
            font=\Large
        ] at (5.5,1.35) {\phantom{$\ket{\psi}=$}};

        %------------------------------------------------
        % Qubits inside the state box
        %------------------------------------------------
        \node[kgfignode, font=\large] (q1) at (-4.2,1.35) {$1$};
        \node[kgfignode, font=\large] (q2) at (-2.5,1.35) {$2$};
        \node at (-1.0,1.35) {$\cdots$};

        \node[kgfignode, font=\large] (v) at (0.7,1.35) {$v$};

        \node at (2.2,1.35) {$\cdots$};
        \node[kgfignode, font=\large] (qn) at (3.9,1.35) {$n$};

    \end{tikzpicture}
\end{figure}
\end{figure}
\vspace{-3mm}
Let $B$ be a basis, and imagine that every qubit $i$ of $\ket{\psi}$ has been measured in $B_i$, save for the qubit $v$. In other words, we measure according to $P = B\setminus \lbrace B_v\rbrace $. Using property \ref{property:stab_no_signs_meas}, the qubit $v$ will be projected into an eigenstate $\ket{R_\pm}$ of some single-qubit Pauli operator $R$. Here $R$ is an element of the skew class $\omega_v =\lbrace X_v, Y_v, Z_v \rbrace $ of $v$, independent of the measurement outcomes on the other qubits.

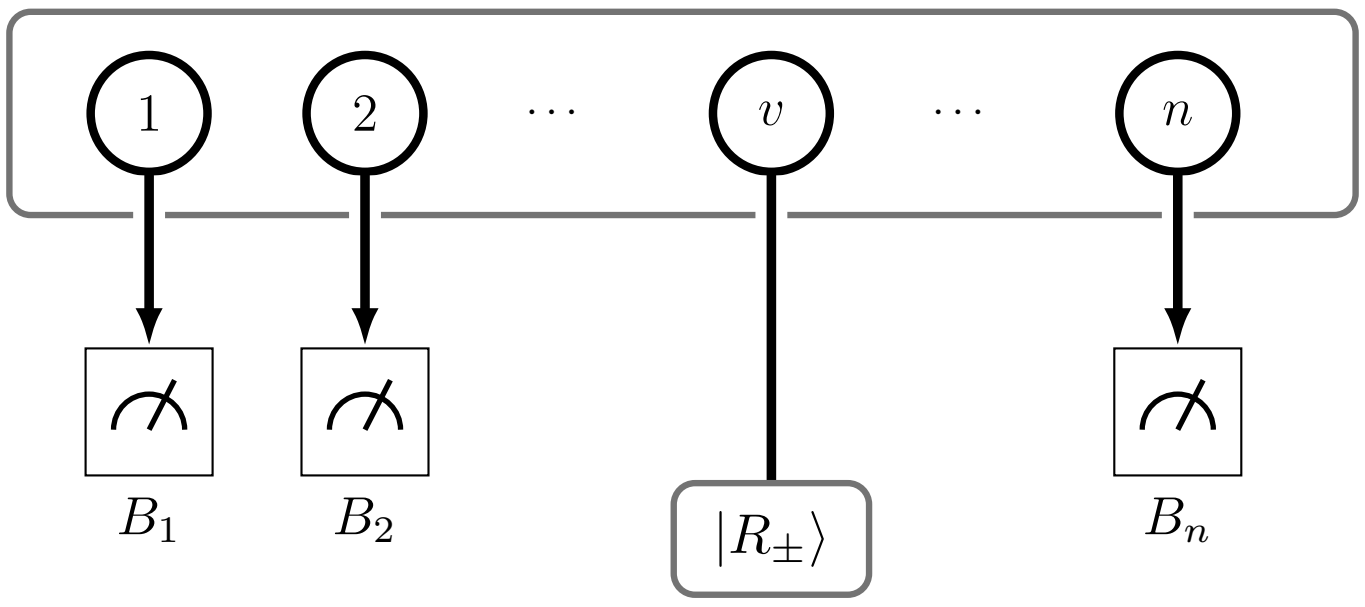
\begin{figure}[H]
    \centerfloat
\begin{figure}[H]
    \centering
    \begin{tikzpicture}[scale=0.84,transform shape]

        %------------------------------------------------
        % Adjustable measurement parameters
        %------------------------------------------------
        \def\meassize{10mm}        % size of measurement box
        \def\measarcradius{2.8mm}  % radius of meter arc
        \def\measneedle{2.5mm}     % length/height of meter needle

        %------------------------------------------------
        % Styles
        %------------------------------------------------
        \tikzset{
            measbox/.style={
                draw,
                minimum width=\meassize,
                minimum height=\meassize,
                inner sep=0pt
            },
            measlabel/.style={
                font=\normalsize
            },
            outcomelabel/.style={
                font=\normalsize
            }
        }

        %------------------------------------------------
        % State box containing the qubits
        %------------------------------------------------
        \node[
            kgfigbox,
            minimum width=10.6cm,
            minimum height=1.6cm
        ] (psi) at (0,1.35) {};

        % \node[scale=2] at (0,2.35) {$\ket{\psi}$};

        %------------------------------------------------
        % Qubits inside the state box
        %------------------------------------------------
        \node[kgfignode, font=\large] (q1) at (-4.2,1.35) {$1$};
        \node[kgfignode, font=\large] (q2) at (-2.5,1.35) {$2$};
        \node at (-1.0,1.35) {$\cdots$};

        \node[kgfignode, font=\large] (v) at (0.7,1.35) {$v$};

        \node at (2.2,1.35) {$\cdots$};
        \node[kgfignode, font=\large    ] (qn) at (3.9,1.35) {$n$};

        %------------------------------------------------
        % Measurements
        %------------------------------------------------
        \node[measbox] (m1) at (-4.2,-1) {};
        \node[measbox] (m2) at (-2.5,-1) {};
        % \node at (-1.0,-0.35) {$\cdots$};

        % \node at (2.2,-0.35) {$\cdots$};
        \node[measbox] (mn) at (3.9,-1) {};

        %------------------------------------------------
        % Measurement icons
        %------------------------------------------------
        \begin{scope}[shift={(m1.center)}]
            \draw[line width=1.0pt]
                (-\measarcradius,-1.4mm)
                arc[start angle=180,end angle=0,radius=\measarcradius];
            \draw[line width=1.0pt]
                (0,-1.4mm) -- (2.0mm,\measneedle);
        \end{scope}

        \begin{scope}[shift={(m2.center)}]
            \draw[line width=1.0pt]
                (-\measarcradius,-1.4mm)
                arc[start angle=180,end angle=0,radius=\measarcradius];
            \draw[line width=1.0pt]
                (0,-1.4mm) -- (2.0mm,\measneedle);
        \end{scope}

        \begin{scope}[shift={(mn.center)}]
            \draw[line width=1.0pt]
                (-\measarcradius,-1.4mm)
                arc[start angle=180,end angle=0,radius=\measarcradius];
            \draw[line width=1.0pt]
                (0,-1.4mm) -- (2.0mm,\measneedle);
        \end{scope}

        %------------------------------------------------
        % Arrows from qubits to measurements
        %------------------------------------------------
        \draw[kgfigarrow] (q1) -- (m1);
        \draw[kgfigarrow] (q2) -- (m2);
        \draw[kgfigarrow] (qn) -- (mn);

        %------------------------------------------------
        % Measurement basis labels
        %------------------------------------------------
        \node[scale=1.2, measlabel] at (-4.2,-1.85) {$B_1$};
        \node[scale=1.2, measlabel] at (-2.5,-1.85) {$B_2$};
        \node[scale=1.2, measlabel] at (3.9,-1.85) {$B_n$};

        %------------------------------------------------
        % Unmeasured qubit v
        %------------------------------------------------
        \draw[kgfigarrow] (v) -- (0.7,-1.85);

        \node[
            scale=1.25,
            kgfigbox,
            inner xsep=7pt,
            inner ysep=5pt
        ] at (0.7,-2) {$\ket{R_{\pm}}$};

    \end{tikzpicture}
\end{figure}
\end{figure}
\vspace{-3mm}
Since the qubit at $v$ is an eigenstate of $R$, measuring $v$ in $R$ will lead to a non-random outcome. As such, $P\cup R$ is not a basis. The converse also holds: measuring qubit $v$ in any of the two remaining Pauli operators $R'$ that is not $R$ will lead to random measurement outcomes on $v$. This means that $P\cup R'$ is a basis as long as $R \neq R'$. One of the two options for $R' \in \omega_v\setminus \lbrace R\rbrace $ is $B_v$. Since we want $B'$ to be different from $B$, the only option is to change $B_v$ to the unique remaining element in $\omega_v\setminus \lbrace B_v, R\rbrace$. \end{proof}
% The above proof relied on a combination of the uncertainty principle (conditioned on previous measurements, there is at most one single-qubit measurement per qubit that can yield a deterministic result) and the fact that the state is pure (measuring all but one qubit leaves this last qubit in a single-qubit stabilizer state). We note here that these two properties are encoded by axiom 2.4 of multimatroids in~\cite{bouchet1998multimatroids}, and the \emph{tightness} condition on multimatroids~\cite{bouchet2001multimatroids}, respectively. As such, 

We now show that switchings really do generalize local complementations.

\begin{theorem}[\cite{bouchet2001multimatroids}]\label{thm:switch_is_LC}
Let $B$ be a basis of a stabilizer state $\ket{\psi}$, let $v$ be an arbitrary qubit of $\ket{\psi}$, and let $G_B$ be the fundamental graph with respect to $B$. Then $G_{\left(B\star v\right)} = G_B \star v$.
\end{theorem}
\begin{proof}
Recall that the closed neighborhood of a vertex $v$ in a fundamental graph $G_B$ is given by $\textrm{supp}(g_v^B)$, where $g_v^B$ is the unique non-identity stabilizer contained in $B\cup \omega_v$ (definitions \ref{def:fundamental_graph} and \ref{def:fund_circuits}). Let $D=B\star v$, and $w$ an arbitrary vertex distinct from $v$. To prove the above theorem, we compare the supports of $g_v^B$ and $g_v^{D}$, and $g_w^B$ and $g_w^{D}$. We split the latter case in two: when $w$ is not adjacent to $v$ and when $w$ is adjacent to $v$. We show that these supports match exactly the closed neighborhoods after locally complementing $G_B$ at $v$, proving the claim.

First, $\textrm{supp}(g_v^B) = \textrm{supp}(g_v^{D})$. This follows from the fact that $B\cup \omega_v = D \cup \omega_v$, since $B$ and $D$ are neighboring at $v$. Second, $\textrm{supp}(g_w^B) = \textrm{supp}(g_w^{D})$ when $w$ is not adjacent to $v$. To see this, note that by definition of the fundamental graph (definition \ref{def:fundamental_graph}), we have that $v\notin \textrm{supp}(g_w^B)$. Since $B$ and $D$ are neighboring at $v$, it follows that $g_w^B$ is also contained in $D\cup \omega_w$, so that $g_w^B=g_w^{D}$, and in particular their supports are the same. We thus find that the closed neighborhoods of $v$ and any $w$ not adjacent to $v$ in $G_{(B\star v)}$ are the same as the closed neighborhoods of $G_B\star v$.

\medskip

The remaining case of $w$ adjacent to $v$ is more complicated. As before, we need to find the support of $g_w^{D}$. We claim that $g_w^{D}= g_v^B\cdot g_w^B$ (where we ignore phases as usual). To prove this, we need to show that $g_v^B\cdot g_w^B$ is a stabilizer contained in $D\cup \omega_w$. Clearly $g_v^B\cdot g_w^B$ is a stabilizer, since it is a product of stabilizers, so it remains to show that $g_v^B\cdot g_w^B\subseteq D\cup \omega_w$. But for every $i\neq v,w$, both $g_v^B$ and $g_w^B$ have either $I_i$ or $B_i$ as their entry at $i$. Their product therefore has either $I_i$ or $B_i$ as its $i$'th entry. At $i=w$ there is no restriction, since the entire skew class $\omega_w$ is contained in $D\cup\omega_w$. It remains only to consider the entry at $v$. The $v$'th entry of $g_v^B$ is the unique element in $\omega_{v}\setminus \lbrace B_v, D_v\rbrace $, while the $v$'th entry of $g_w^{B}$ is $B_v$. As such, their product equals $D_v$, so that indeed $g_v^B\cdot g_w^B\subseteq D\cup \omega_w$. Thus, by the uniqueness of fundamental circuits, we have that $g_w^D=g_v^B\cdot g_w^B$.

The fundamental circuit $g_w^{D}$ has the desired support of
\begin{align}
\textrm{supp}\left(g_w^{D}\right)=\left(\textrm{supp}\left(g_v^{B}\right)\Delta  \, \textrm{supp}\left(g_w^{B}\right)\right)\, \cup\,  \lbrace v, w\rbrace \ ,\label{eq:supp_cond}
\end{align}
since the closed neighborhood of $w$ after complementing the graph $G_B$ at $v$ equals $N_v\Delta N_w$. The equality in Eq.~\eqref{eq:supp_cond} holds since both $g_v^B$ and $g_w^B$ have either $I_i$ or $B_i$ for their $i$'th entry, as long as $i\neq v, w$. For $i\in \lbrace v, w\rbrace $, the entries of $g_v^B$ and $g_w^B$ at $i$ are two distinct non-identity Paulis, so that their product is a non-identity Pauli as well. \end{proof}

\subsection{So what are local complementations, really?}
We now briefly summarize how graphs up to local complementations are better thought of as bases up to switchings. Bases are precisely the local Pauli measurements on all qubits that yield completely random outcomes on a given state $\ket{\psi}$ (property \ref{prop:bases}). For each such basis $B$, there is an associated fundamental graph $G_B$ (definition \ref{def:fundamental_graph}), such that $\ket{\psi}\simeq_{\textrm{LC}}\ket{G_B}$, and every graph state $\ket{H}$ locally equivalent to $\ket{\psi}$ satisfies $H=G_B$ for some $B$ (theorem~\ref{theorem:bases_surj_into_graphs}). Finally, switchings on bases (definition \ref{def:switch_basis}) generalize local complementations, in the sense that $G_{(B\star v)}= G_B \star v$.

Switching a basis $B$ at a vertex $v$ has a concrete physical interpretation: measuring according to $B\setminus \lbrace B_v\rbrace $ leaves qubit $v$ in an eigenstate of a Pauli operator $R$. Of the three Paulis at $v$, measuring in $R$ would yield a deterministic outcome, while the other two measurements yield random outcomes. Switching $B$ at $v$ replaces $B_v$ by the unique other Pauli that is not equal to $R$.

% To a physicist, a local complementation corresponds to a local Clifford operation $\sqrt{X_vZ_{N_v}}$ mapping one graph state to another. Switchings on bases are very different, in the sense that they are just a change of representation, but for a fixed state. 

The fundamental graphs $G_B$ also carry a concrete physical interpretation. Measuring each qubit according to $B$ yields random measurement outcomes. By changing the measurement $B_v$ at a qubit $v$ appropriately, the outcomes are no longer random. More specifically, correlations will now be observed among the measurement outcomes on the support of the fundamental circuit $g_v^B$. The fact that $g_v^B$ is the unique circuit contained in $B\cup \omega_v$ implies that these are the only correlations observed. 

Note how the fundamental graphs and bases are properties of a fixed stabilizer state $\ket{\psi}$. This is quite different from how local complementations are normally interpreted in the quantum info literature. Namely, a local complementation at $v$ corresponds to a Clifford unitary $\sqrt{X_v}\prod_{w\in N_v}\sqrt{Z_{w}}$, mapping $\ket{G}$ to $\ket{G\star v}$. Arguably the perspective where the state is fixed and one considers different representations is more natural than actively changing the state. We provide an interactive visualization of the above ideas in~\cite{visualization}, which---perhaps not surprisingly---was created entirely by ChatGPT.

In the next section, we will see what the vantage point above gives us. 
\newpage
\section{What use are bases?}\label{sec:use_of_bases}
\subsection{Bases as a proof technique}
At this point, it is natural to wonder what switchings on bases buy us that we cannot get from local complementations on graphs. Indeed, in many settings, working with bases is more natural than working with graphs. For example, bases have been used in~\cite{dahlberg2022complexity} to prove that the so-called vertex-minor problem is NP-complete, in~\cite{dahlberg2020counting} to prove that counting the number of graphs locally equivalent to a given graph is $\#$P-complete, and in~\cite{bahramgiri2007enumerating} to understand local equivalence for the qudit case\footnote{Note that these papers refer to bases as \emph{Eulerian vectors}.}.

We now provide three statements on graphs up to local complementations that are more easily proved using bases.

\begin{theorem}[\cite{bahramgiri2007enumerating}]
For an arbitrary graph $G$ on $n$ vertices, there are at most $3^n$ graphs locally equivalent to $G$.
\end{theorem}
\begin{proof}
Theorem \ref{theorem:bases_surj_into_graphs} tells us that for each graph $H$ locally equivalent to $G$, there is a basis $B$ of $\ket{G}$ such that $G_B = H$. This implies that there are at least as many bases of $\ket{G}$ as there are graphs locally equivalent to $G$. Thus, to upper bound the number of elements in the equivalence class, it suffices to upper bound the number of bases.

Here is a seemingly `cheap' such upper bound: since bases need to be full-weight Pauli strings, clearly there can be no more bases than there are full-weight Pauli strings, i.e.~$\left|\mathcal{B}\right| \leq 3^n$. 
\end{proof}
The above bound might sound loose, but random graph states achieve $\left|\mathcal{B}\right| \sim c\cdot3^n$ for some explicitly known constant $c$! We prove this result in follow-up work. 

How about bounding the number of local complementations needed to go from $G$ to $H$ when $G$ and $H$ are locally equivalent?

\begin{theorem}[Upper bound on number of local complementations~\cite{bouchet1990kappa}]\label{thm:diameter}
If $G$ and $H$ are locally equivalent graphs on $n$ vertices, then at most $\frac{3}{2}n$ local complementations are needed to transform $G$ into $H$.
\end{theorem}
\begin{proof}
We provide here a sketch of the proof, see theorem \ref{thm:diameter2} in the appendix for more details.

It suffices to prove that we need at most $\frac{3}{2}n$ switchings to map any basis $B$ to another basis $B'$. We first perform switchings at vertices $v$ that take $B_v$ to $B_v'$. At a certain point, no such vertices exist anymore. We then show that so-called pivots~\cite{oum2023rank, bouchet2001multimatroids} on certain pairs of vertices $v, w$ allow us to map both $B_v$ and $B_w$ to $B_v'$ and $B_w'$, respectively. Such pivots are implemented by switching on $v$, $w$ and again on $v$, and thus allow us to map two entries of $B$ to two entries of $B'$. Since three switches `fix' two entries, we find an upper bound of $\frac{3}{2}n$. The technical part of the proof involves showing that such pairs $v, w$ always exist, i.e.~that one cannot get `stuck' during this process.
\end{proof}

While the bound in theorem \ref{thm:diameter} achieves the right scaling, it is not tight. See reference~\cite{fon1989distance} for the stronger upper bound of $\max\left(n+1, \frac{10}{9}n\right)$ (see also Ref.~\cite{bouchet1990kappa}).

Recall from section \ref{sec:fundamental_graphs} that a circuit is a Pauli string $P$ such that the only stabilizers in $P$ are the identity and $P$ itself. The following theorem on circuits is surprisingly messy to prove without using the concept of a basis, see for example Ref.~\cite{goodenough2026optimal}.

\begin{theorem}[Circuits to canonical generators of graphs~\cite{brijder2015isotropic1}]\label{thm:circ_to_canonical_gens}
Let $P$ be a circuit of a stabilizer state $\ket{\psi}$, and let $v\in \textrm{supp}(P)$ be arbitrary. Then $\ket{\psi}$ is local-Clifford equivalent to a graph state $\ket{G}$ such that the neighborhood of $v$ in the graph $G$ equals $\textrm{supp}(P)\setminus\lbrace v\rbrace$.
\end{theorem}

To prove it, we use the following lemma.

\begin{lemma}\label{lemma:extend_to_basis}
Let $\ket{\psi}$ be a stabilizer state, and $Q$ an arbitrary (not necessarily full-weight) Pauli string such that measuring according to $Q$ yields $2^{\left|\textrm{supp}(Q)\right|}$ random outcomes. Then there exists a basis $B\in \mathcal{B}_\psi$ such that $Q\subseteq B$.
\end{lemma}
\begin{proof}
See lemma \ref{lemma:extend_to_basis2} in the appendix.
\end{proof}

\begin{proof}[Proof of theorem \ref{thm:circ_to_canonical_gens}]
Consider the case where each qubit $i \in \textrm{supp}(P) \setminus \lbrace v\rbrace$ is measured according to $P_i$. Since $P$ is a stabilizer, qubit $v$ will be left in an eigenstate of $P_v$ (see property \ref{property:stab_no_signs_meas}).

Now change $P_v$ to one of the two non-identity Paulis $P_v'$, yielding another Pauli string $Q:=P\Delta \lbrace P_v, P_v'\rbrace$. Since $P$ is a circuit, measuring the qubits according to $Q$ yields completely random outcomes. Using lemma \ref{lemma:extend_to_basis}, we can extend $Q$ to a basis $B$, i.e.~there exists $B\in \mathcal{B}_\psi$ such that $Q\subseteq B$.

By construction, the neighborhood of the vertex $v$ in the fundamental graph $G_B$ equals $\textrm{supp}(P)\setminus\lbrace v\rbrace$. To see this, note that $B\cup \omega_v$ contains the stabilizer $P$, so that $g_v^B=P$ by definition \ref{def:fund_circuits}. Thus by definition \ref{def:fundamental_graph}, $\ket{\psi}$ has a basis $B$ such that the fundamental graph $G_B$ has the desired closed neighborhood of $v$. From theorem~\ref{theorem:bases_surj_into_graphs} we have then that $\ket{\psi}\simeq_{\textrm{LC}}\ket{G_B}$, proving the claim.
\end{proof}

\subsection{Switching graphs}
The set of bases $\mathcal{B}_\psi$ of a state $\ket{\psi}$ carries more than just a set structure. Namely, two bases $B$ and $B'$ are `close' if they differ by a single switching. This motivates the definition of a \emph{switching graph}~\cite{bouchet1990kappa}.

\begin{definition}[Switching graphs of stabilizer states]
The switching graph $\mathcal{S}(\ket{\psi})$ of a stabilizer state $\ket{\psi}$ is the graph with vertex set $\mathcal{B}_\psi$, and where two vertices $B, B'$ are adjacent $\iff$ $B$ and $B'$ are related by a single switching.
\end{definition}

\begin{example}\label{example:basis_graphs}
The switching graphs of $\ket{00}$ and $\ket{00}+\ket{11}$ are the $4$- and $6$-cycles seen below. 
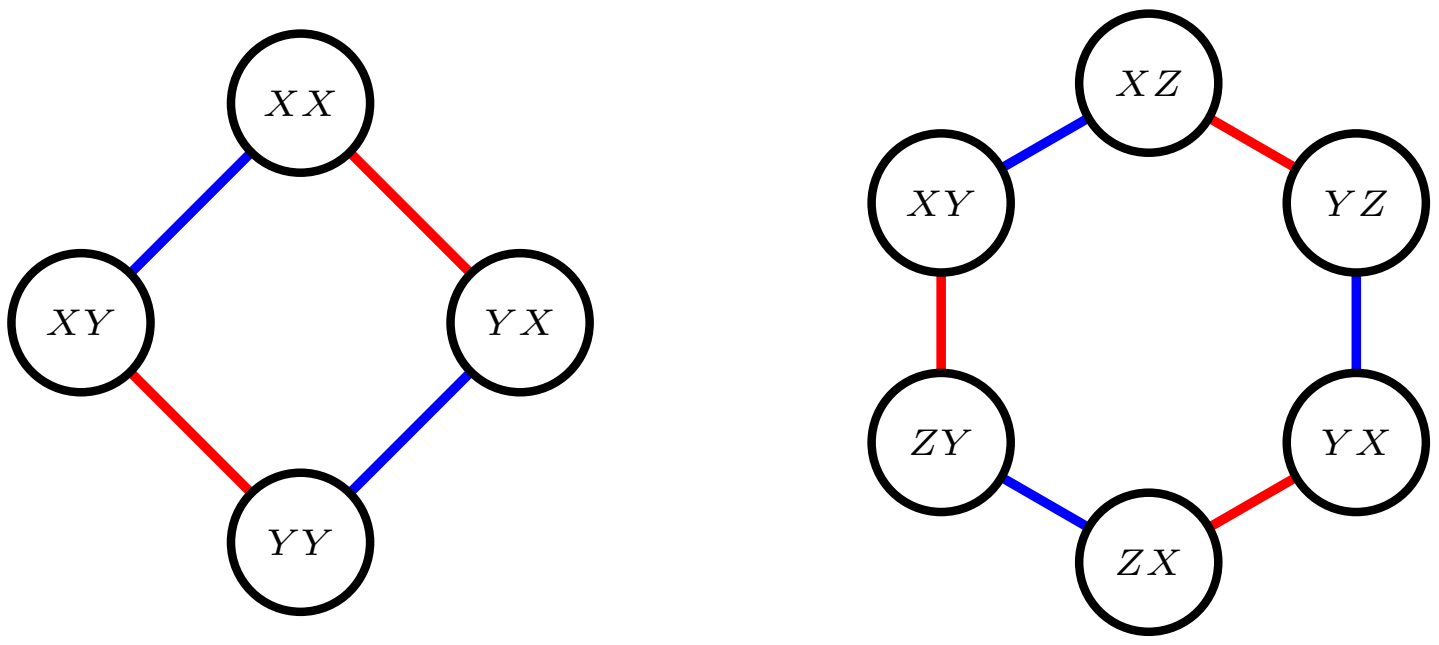
\begin{figure}[H]
    \centerfloat
\begin{tikzpicture}[
    scale=0.66,
    every node/.style={font=\Large},
    edge/.style={
        line width=1.8pt,
        line cap=round,
        line join=round,
        preaction={draw=white, line width=6pt}
    },
    qone/.style={edge, draw=red},
    qtwo/.style={edge, draw=blue},
    hubanchor/.style={
        circle,
        minimum size=9.2mm,
        inner sep=0pt
    },
    hub/.style={
        circle,
        draw,
        fill=white,
        fill opacity=1,
        text opacity=1,
        line width=1.6pt,
        minimum size=9.2mm,
        inner sep=0pt
    },
    vlabel/.style={font=\scriptsize}
]

    %%%%%%%%%%%%%%%%%%%%%%%%%
    % Left figure
    %%%%%%%%%%%%%%%%%%%%%%%%%
    \begin{scope}

        % Coordinates
        \coordinate (ca) at (0,2.2);
        \coordinate (cb) at (2.2,0);
        \coordinate (cc) at (0,-2.2);
        \coordinate (cd) at (-2.2,0);

        % Invisible circular anchors
        \node[hubanchor] (a) at (ca) {};
        \node[hubanchor] (b) at (cb) {};
        \node[hubanchor] (c) at (cc) {};
        \node[hubanchor] (d) at (cd) {};

        % Exchange edges
        \draw[qone] (a) -- (b);
        \draw[qtwo] (b) -- (c);
        \draw[qone] (c) -- (d);
        \draw[qtwo] (d) -- (a);

        % Visible vertices and labels
        \node[hub] at (ca) {};
        \node[vlabel] at (ca) {$XX$};

        \node[hub] at (cb) {};
        \node[vlabel] at (cb) {$YX$};

        \node[hub] at (cc) {};
        \node[vlabel] at (cc) {$YY$};

        \node[hub] at (cd) {};
        \node[vlabel] at (cd) {$XY$};

    \end{scope}

    %%%%%%%%%%%%%%%%%%%%%%%%%
    % Right figure
    %%%%%%%%%%%%%%%%%%%%%%%%%
    \begin{scope}[xshift=8.5cm]

        % Coordinates
        \coordinate (cu1) at (0,2.4);
        \coordinate (cu2) at (2.08,1.2);
        \coordinate (cu3) at (2.08,-1.2);
        \coordinate (cu4) at (0,-2.4);
        \coordinate (cu5) at (-2.08,-1.2);
        \coordinate (cu6) at (-2.08,1.2);

        % Invisible circular anchors
        \node[hubanchor] (u1) at (cu1) {};
        \node[hubanchor] (u2) at (cu2) {};
        \node[hubanchor] (u3) at (cu3) {};
        \node[hubanchor] (u4) at (cu4) {};
        \node[hubanchor] (u5) at (cu5) {};
        \node[hubanchor] (u6) at (cu6) {};

        % Exchange edges
        \draw[qone] (u1) -- (u2);
        \draw[qtwo] (u2) -- (u3);
        \draw[qone] (u3) -- (u4);
        \draw[qtwo] (u4) -- (u5);
        \draw[qone] (u5) -- (u6);
        \draw[qtwo] (u6) -- (u1);

        % Visible vertices and labels
        \node[hub] at (cu1) {};
        \node[vlabel] at (cu1) {$XZ$};

        \node[hub] at (cu2) {};
        \node[vlabel] at (cu2) {$YZ$};

        \node[hub] at (cu3) {};
        \node[vlabel] at (cu3) {$YX$};

        \node[hub] at (cu4) {};
        \node[vlabel] at (cu4) {$ZX$};

        \node[hub] at (cu5) {};
        \node[vlabel] at (cu5) {$ZY$};

        \node[hub] at (cu6) {};
        \node[vlabel] at (cu6) {$XY$};

    \end{scope}

\end{tikzpicture}
\end{figure}
The switchings at qubits $1$ and $2$ are labelled by red and blue edges, respectively. E.g., performing a switching at qubit $1$ on $B=YX$ yields $XX$ for the $\ket{00}$ state; switching at qubit $2$ on $B=YZ$ yields $YX$ for $\ket{00}+\ket{11}$.
\end{example}

\begin{example}
The switching graph of the separable stabilizer state $\ket{0}^{\otimes n}$ is isomorphic to the hypercube graph, since its bases are all possible strings consisting of $X$ and $Y$ operators.
\end{example}

\begin{example}
The switching graph of the complete graph state $\ket{K_3}$ on three qubits is as follows.
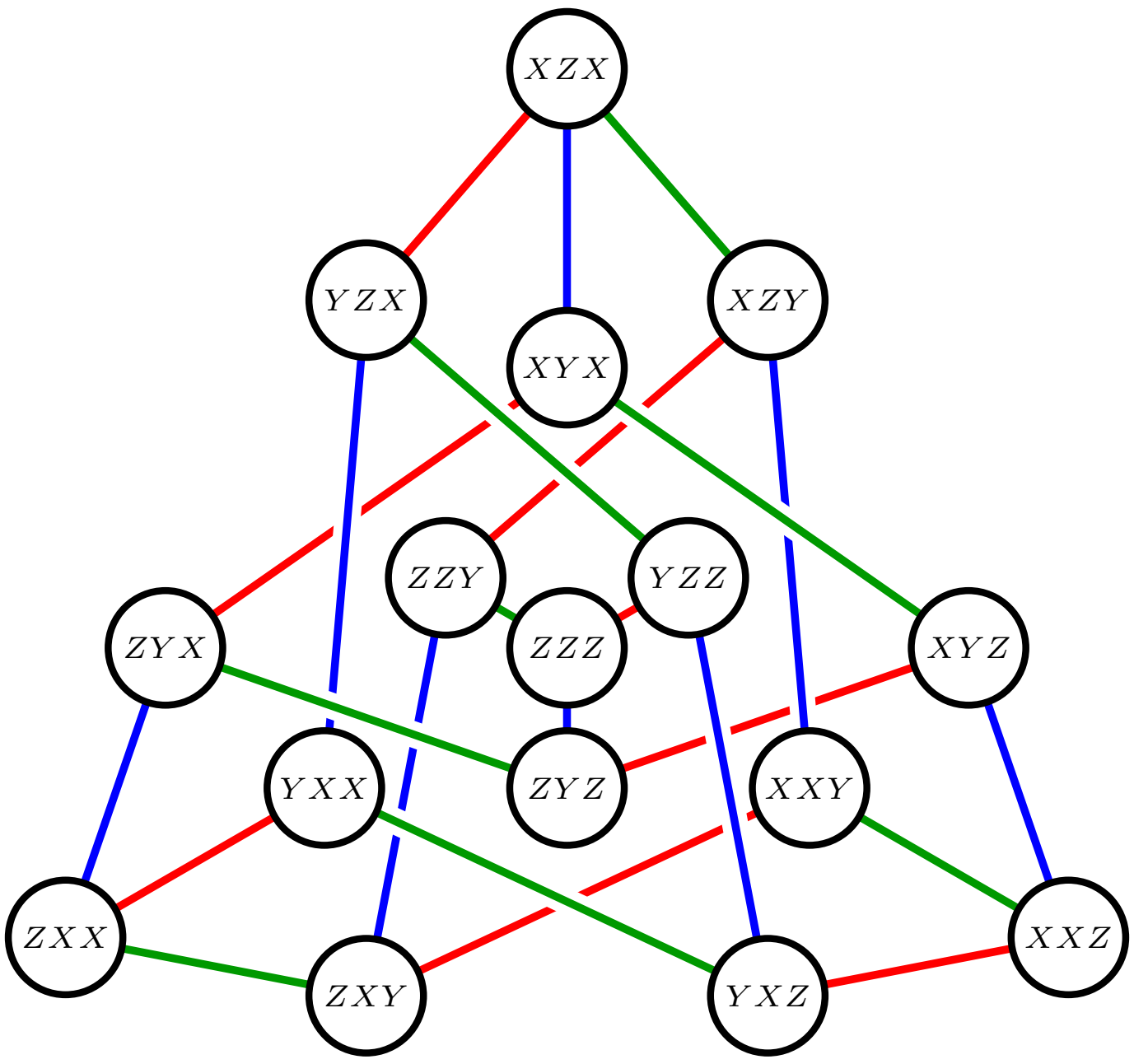
\begin{figure}[H]
    \centering
\begin{tikzpicture}[
    scale=0.75,
    every node/.style={font=\Large},
    edge/.style={
        line width=1.8pt,
        line cap=round,
        line join=round,
        preaction={draw=white, line width=6pt}
    },
    qone/.style={edge, draw=red},
    qtwo/.style={edge, draw=blue},
    qthree/.style={edge, draw=green!60!black},
    hubanchor/.style={
        circle,
        minimum size=9.2mm,
        inner sep=0pt
    },
    hub/.style={
        circle,
        draw,
        fill=white,
        fill opacity=1,
        text opacity=1,
        line width=1.6pt,
        minimum size=9.2mm,
        inner sep=0pt
    },
    vlabel/.style={font=\scriptsize}
]

    %--------------------------------------------------
    % Symmetry convention
    %--------------------------------------------------
    % A 120-degree rotation sends
    %
    %     P_1 P_2 P_3  ->  P_2 P_3 P_1.
    %
    % Reflection in the vertical axis sends
    %
    %     P_1 P_2 P_3  ->  P_3 P_2 P_1.
    %
    % Thus the underlying graph has D_3 symmetry.
    % With edge colors, reflection swaps qone <-> qthree
    % and fixes qtwo.

    \def\Rhex{4.30}
    \def\Rout{6.20}
    \def\Rmid{3.00}
    \def\Rin{1.50}

    \newcommand{\placebasis}[3]{%
        \coordinate (c#1) at ({#2*cos(#3)}, {#2*sin(#3)});
    }

    %--------------------------------------------------
    % Coordinates
    %--------------------------------------------------

    % Six-element orbit:
    % {XYZ, YZX, ZXY, ZYX, YXZ, XZY}.
    %
    % These are placed on a regular hexagon.
    \placebasis{XYZ}{\Rhex}{0}
    \placebasis{XZY}{\Rhex}{60}
    \placebasis{YZX}{\Rhex}{120}
    \placebasis{ZYX}{\Rhex}{180}
    \placebasis{ZXY}{\Rhex}{240}
    \placebasis{YXZ}{\Rhex}{300}

    % Outer triangle:
    % {XXZ, XZX, ZXX}.
    \placebasis{XZX}{\Rout}{90}
    \placebasis{ZXX}{\Rout}{210}
    \placebasis{XXZ}{\Rout}{330}

    % Middle triangle:
    % {XXY, XYX, YXX}.
    \placebasis{XYX}{\Rmid}{90}
    \placebasis{YXX}{\Rmid}{210}
    \placebasis{XXY}{\Rmid}{330}

    % Inner inverted triangle:
    % {YZZ, ZZY, ZYZ}.
    \placebasis{ZYZ}{\Rin}{270}
    \placebasis{YZZ}{\Rin}{30}
    \placebasis{ZZY}{\Rin}{150}

    % Fixed point.
    \coordinate (cZZZ) at (0,0);

    %--------------------------------------------------
    % Invisible circular anchors
    %--------------------------------------------------
    \def\allbases{%
        XXY,XXZ,XYX,XYZ,%
        XZX,XZY,YXX,YXZ,%
        YZX,YZZ,ZXX,ZXY,%
        ZYX,ZYZ,ZZY,ZZZ%
    }

    \foreach \v in \allbases {
        \node[hubanchor] (\v) at (c\v) {};
    }

    %--------------------------------------------------
    % Exchange edges
    %--------------------------------------------------

    % qubit 1, red
    \draw[qone] (XXY) -- (ZXY);
    \draw[qone] (XXZ) -- (YXZ);
    \draw[qone] (XYX) -- (ZYX);
    \draw[qone] (XYZ) -- (ZYZ);
    \draw[qone] (XZX) -- (YZX);
    \draw[qone] (XZY) -- (ZZY);
    \draw[qone] (YXX) -- (ZXX);
    \draw[qone] (YZZ) -- (ZZZ);

    % qubit 2, blue
    \draw[qtwo] (XXY) -- (XZY);
    \draw[qtwo] (XXZ) -- (XYZ);
    \draw[qtwo] (XYX) -- (XZX);
    \draw[qtwo] (YXX) -- (YZX);
    \draw[qtwo] (YXZ) -- (YZZ);
    \draw[qtwo] (ZXX) -- (ZYX);
    \draw[qtwo] (ZXY) -- (ZZY);
    \draw[qtwo] (ZYZ) -- (ZZZ);

    % qubit 3, green
    \draw[qthree] (XXY) -- (XXZ);
    \draw[qthree] (XYX) -- (XYZ);
    \draw[qthree] (XZX) -- (XZY);
    \draw[qthree] (YXX) -- (YXZ);
    \draw[qthree] (YZX) -- (YZZ);
    \draw[qthree] (ZXX) -- (ZXY);
    \draw[qthree] (ZYX) -- (ZYZ);
    \draw[qthree] (ZZY) -- (ZZZ);

    %--------------------------------------------------
    % Visible vertices and labels, drawn last
    %--------------------------------------------------
    \foreach \v in \allbases {
        \node[hub] at (c\v) {};
        \node[vlabel] at (c\v) {$\v$};
    }

\end{tikzpicture}
\end{figure}
\vspace{-4mm}
Here red corresponds to switching on the first qubit, and blue and green correspond to the second and third qubit, respectively.
\end{example}

Since local Cliffords only relabel the vertices, the switching graphs (up to isomorphism) are an \emph{invariant} under local Cliffords. In fact, something stronger is true: the bases (and thus the labelled switching graph as well) specify a stabilizer state uniquely up to Pauli unitaries, see~\cite{bouchet1997multimatroids}.

The reader may have noticed the similarities between the switching graph and \emph{LC-orbits} of graphs/graph states~\cite{adcock2020mapping}. Here, the LC-orbit of $G$ is itself a graph $\mathcal{LC}(\ket{G})$, whose vertex set equals the whole equivalence class of $G$, and where $H$ and $K$ are adjacent $\iff$ $H$ and $K$ are related by a single local complementation~\cite{adcock2020mapping}.

LC-orbits have been studied in \cite{adcock2020mapping, cabello2013exclusivity, burchardt2025foliage,revis2025orbit, danielsen2010interlace, genestcomplementation}, where in particular attempts have been made to connect entanglement properties of $\ket{G}$ with properties of the graph $\mathcal{LC}(\ket{G})$. It is however easy to see that $\mathcal{LC}(\ket{G})$ cannot say much about entanglement of $\ket{G}$ itself. This is because both $\ket{\graphNoEdge}:=\ket{++}$ and $\ket{\graphEdge}:=CZ\ket{++}$ are invariant under local complementations, meaning that both $\mathcal{LC}(\ket{\graphNoEdge})$ and $\mathcal{LC}(\ket{\graphEdge})$ are trivial graphs! In contrast, the switching graphs $\mathcal{S}(\ket{\graphNoEdge})$ and $\mathcal{S}(\ket{\graphEdge})$ are isomorphic to the switching graphs from example \ref{example:basis_graphs}, and thus \emph{can} distinguish the two states (and their entanglement) from one another. 

Let us briefly detail why the LC-orbit loses information that is contained in the switching graph. From subsection \ref{sec:fund_graphs_of_bases}, every basis $B$ has an associated fundamental graph $G_B$. In certain cases, different bases $B, B'$ have the same fundamental graph, i.e.~$G_B=G_{B'}$. The states $\ket{00}$ and $\ket{00}+\ket{11}$ are extreme examples of this, where every basis has the same fundamental graph.

This phenomenon was already observed by Bouchet in~\cite{bouchet1993recognizing}, where he noted that the preimage of any $G_B$ under the map $B\mapsto G_B$ has the same cardinality $k$, independent of $G_B$. This implies that $\frac{\left|\mathcal{B}_G\right|}{\left|\mathcal{LC}(\ket{G})\right|}=k$. Bouchet describes in \cite{bouchet1993recognizing} a technique to calculate the quantity $k$, which he calls the \emph{index}. 
These observations were used in \cite{dahlberg2020counting} to prove that counting the number of locally equivalent graphs is $\#$P-complete.

Bouchet expresses the computation of $k$ in terms of the so-called bineighborhood of graphs~\cite{bouchet1993recognizing}. There is, however, a simple conceptual interpretation of the parameter $k$ in terms of quantum information. Two bases $B, B'\in \mathcal{B}_\psi$  map to the same fundamental graph $G_B=G_{B'}$ if and only if there is a local Clifford symmetry $C=\otimes_{v\in V}\,  C_v$ of $\ket{\psi}$ such that $B'=CBC^\dagger$\footnote{We do not consider phases here, so local Clifford symmetries \emph{up to phases and Paulis} is more accurate.}. As such, $k$ is equal to the cardinality of the group of local Clifford symmetries of $\ket{\psi}$. We provide two examples of this fact, and leave a proof to the appendix, see theorem~\ref{thm:k_is_sym}.

\begin{example}
We have that $\mathcal{S}(\ket{\graphNoEdge})$ and $\mathcal{LC}(\ket{\graphNoEdge})$ have order $4$ and $1$, so $k$ needs to equal $4$. Indeed, the state $\ket{++}$ is invariant under $\langle HSH\otimes I, I\otimes HSH\rangle$, which has order $4$ up to Paulis/phases. Similarly, $\mathcal{S}(\ket{\graphEdge})$ and $\mathcal{LC}(\ket{\graphEdge})$ have order $6$ and $1$, so $k$ should equal $6$. This matches the symmetries of the state $\ket{00}+\ket{11}$, since it is invariant under any Clifford unitary of the form $C\otimes C^*$. Since phaseless single-qubit Cliffords permute only $X, Y$ and $Z$, they have the same order as the permutation group on three elements, which equals $6$.
\end{example}

\begin{example}
For the complete graph state $\ket{K_3}$ on three qubits, we have that $\mathcal{S}(\ket{K_3})$ and $\mathcal{LC}(\ket{K_3})$ have order $16$ and $4$ (see~\cite{adcock2020mapping}), so the index needs to equal $k=4$. Indeed, the three-qubit GHZ state (which is locally equivalent to $\ket{K_3}$) has local Clifford symmetry group $\langle S\otimes S \otimes I, I \otimes S \otimes S\rangle $, which has order $4$~\cite{wang2026efficient}. We stress once more that we are considering only symmetries up to Paulis. Below we show $\mathcal{S}(\ket{K_3})$ and $\mathcal{LC}(\ket{K_3})$, where red, blue, green correspond to switching/locally complementing on qubit $1$, $2$, $3$, respectively. We also colored the vertices of both graphs to indicate the associated (fundamental) graphs, e.g.~the yellow basis $ZZZ$ has $K_3$ as its fundamental graph, and both $YZZ$ and $YXZ$ have the star graph with $1$ as its center as their associated fundamental graph. Note how the LC-orbit $\mathcal{LC}(\ket{K_3})$ `sits inside' of $\mathcal{S}(\ket{K_3})$.

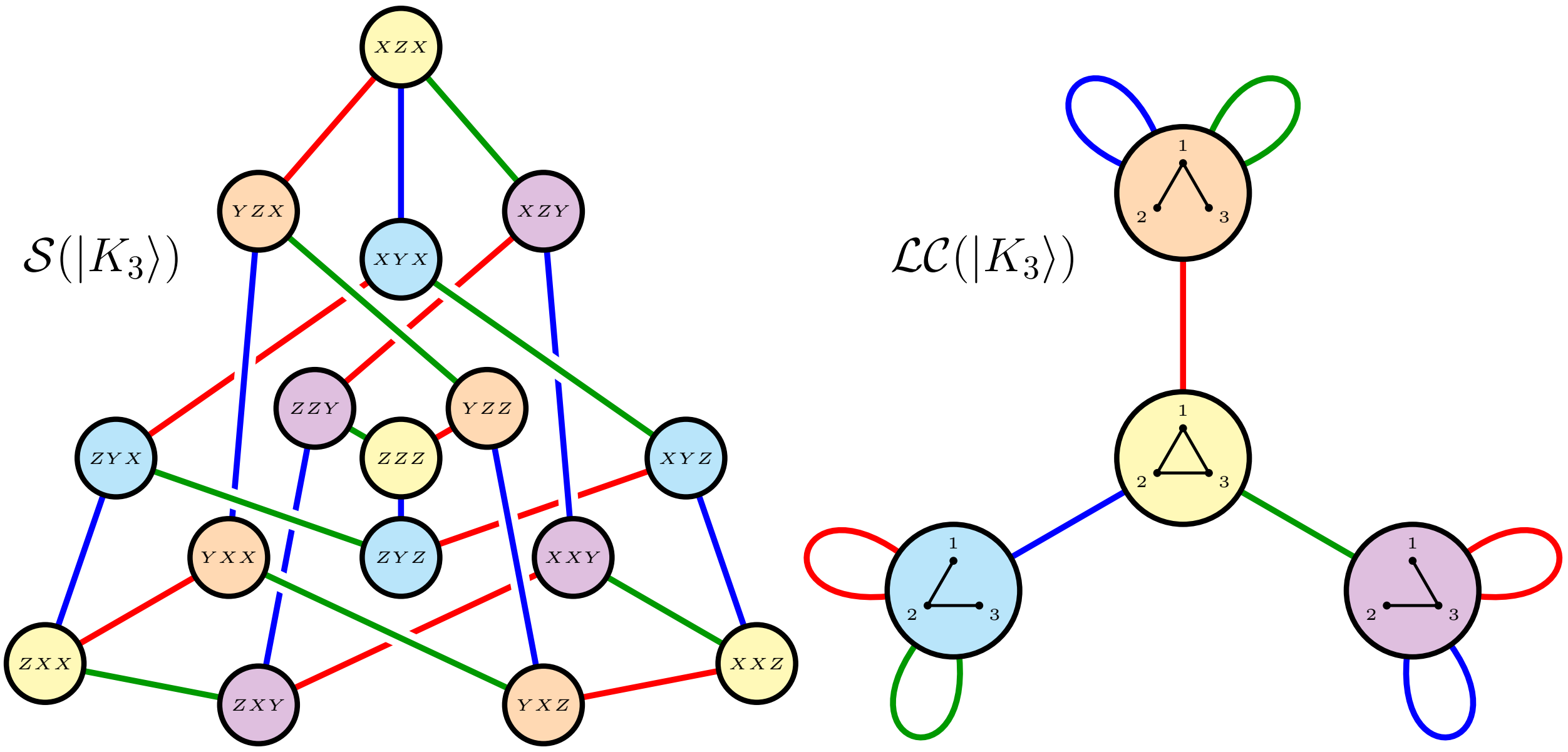
\begin{figure}[H]
    \centering
% Requires \usepackage{tikz} in the preamble.

\begin{tikzpicture}[
    scale=0.7,
    every node/.style={font=\Large},
    edge/.style={
        line width=1.8pt,
        line cap=round,
        line join=round,
        preaction={draw=white, line width=6pt}
    },
    qone/.style={edge, draw=red},
    qtwo/.style={edge, draw=blue},
    qthree/.style={edge, draw=green!60!black},
    hubanchor/.style={
        circle,
        minimum size=8.2mm,
        inner sep=0pt
    },
    hub/.style={
        hubanchor,
        draw,
        fill=white,
        fill opacity=1,
        text opacity=1,
        line width=1.6pt
    },
    vlabel/.style={font=\tiny},
    orbitanchor/.style={
        hubanchor,
        minimum size=14mm
    },
    orbitloop/.style={
        loop,
        looseness=7,
        min distance=11mm
    }
]

    %--------------------------------------------------
    % Parameters
    %--------------------------------------------------

    \def\Rhex{4.30}
    \def\Rout{6.20}
    \def\Rmid{3.00}
    \def\Rin{1.50}

    % Position and radius of the LC orbit.
    \def\OrbitX{11.8}
    \def\OrbitR{4.0}

    % Size of the miniature graphs and label positions.
    \def\MiniGraphR{0.45}
    \def\MiniLabelR{0.72}

    \newcommand{\placebasis}[3]{%
        \coordinate (c#1) at ({#2*cos(#3)}, {#2*sin(#3)});
    }

    %--------------------------------------------------
    % Basis coordinates
    %--------------------------------------------------

    % Regular hexagon.
    \placebasis{XYZ}{\Rhex}{0}
    \placebasis{XZY}{\Rhex}{60}
    \placebasis{YZX}{\Rhex}{120}
    \placebasis{ZYX}{\Rhex}{180}
    \placebasis{ZXY}{\Rhex}{240}
    \placebasis{YXZ}{\Rhex}{300}

    % Outer triangle.
    \placebasis{XZX}{\Rout}{90}
    \placebasis{ZXX}{\Rout}{210}
    \placebasis{XXZ}{\Rout}{330}

    % Middle triangle.
    \placebasis{XYX}{\Rmid}{90}
    \placebasis{YXX}{\Rmid}{210}
    \placebasis{XXY}{\Rmid}{330}

    % Inner inverted triangle.
    \placebasis{ZYZ}{\Rin}{270}
    \placebasis{YZZ}{\Rin}{30}
    \placebasis{ZZY}{\Rin}{150}

    % Centre.
    \coordinate (cZZZ) at (0,0);

    %--------------------------------------------------
    % Invisible basis anchors
    %--------------------------------------------------

    \def\allbases{%
        XXY,XXZ,XYX,XYZ,%
        XZX,XZY,YXX,YXZ,%
        YZX,YZZ,ZXX,ZXY,%
        ZYX,ZYZ,ZZY,ZZZ%
    }

    \foreach \v in \allbases {
        \node[hubanchor] (\v) at (c\v) {};
    }

    %--------------------------------------------------
    % LC-orbit coordinates and invisible anchors
    %
    % K  = triangle.
    % Pi = path centred at qubit i.
    %--------------------------------------------------

    \coordinate (cK) at (\OrbitX,0);

    \coordinate (cP1) at
        ({\OrbitX+\OrbitR*cos(90)},
         {\OrbitR*sin(90)});

    \coordinate (cP2) at
        ({\OrbitX+\OrbitR*cos(210)},
         {\OrbitR*sin(210)});

    \coordinate (cP3) at
        ({\OrbitX+\OrbitR*cos(330)},
         {\OrbitR*sin(330)});

    \foreach \g in {K,P1,P2,P3} {
        \node[orbitanchor] (\g) at (c\g) {};
    }

    %--------------------------------------------------
    % Exchange edges on the left
    %--------------------------------------------------

    % Qubit 1, red.
    \draw[qone] (XXY) -- (ZXY);
    \draw[qone] (XXZ) -- (YXZ);
    \draw[qone] (XYX) -- (ZYX);
    \draw[qone] (XYZ) -- (ZYZ);
    \draw[qone] (XZX) -- (YZX);
    \draw[qone] (XZY) -- (ZZY);
    \draw[qone] (YXX) -- (ZXX);
    \draw[qone] (YZZ) -- (ZZZ);

    % Qubit 2, blue.
    \draw[qtwo] (XXY) -- (XZY);
    \draw[qtwo] (XXZ) -- (XYZ);
    \draw[qtwo] (XYX) -- (XZX);
    \draw[qtwo] (YXX) -- (YZX);
    \draw[qtwo] (YXZ) -- (YZZ);
    \draw[qtwo] (ZXX) -- (ZYX);
    \draw[qtwo] (ZXY) -- (ZZY);
    \draw[qtwo] (ZYZ) -- (ZZZ);

    % Qubit 3, green.
    \draw[qthree] (XXY) -- (XXZ);
    \draw[qthree] (XYX) -- (XYZ);
    \draw[qthree] (XZX) -- (XZY);
    \draw[qthree] (YXX) -- (YXZ);
    \draw[qthree] (YZX) -- (YZZ);
    \draw[qthree] (ZXX) -- (ZXY);
    \draw[qthree] (ZYX) -- (ZYZ);
    \draw[qthree] (ZZY) -- (ZZZ);

    %--------------------------------------------------
    % LC edges on the right
    %--------------------------------------------------

    \draw[qone]   (K) -- (P1);
    \draw[qtwo]   (K) -- (P2);
    \draw[qthree] (K) -- (P3);

    %--------------------------------------------------
    % Self-loops
    %
    % The loop geometry is related by 120-degree
    % rotations at the three outer vertices.
    %--------------------------------------------------

    % P1: leaves 2 and 3.
    \draw[qtwo]
        (P1) edge[orbitloop,out=155,in=115] (P1);
    \draw[qthree]
        (P1) edge[orbitloop,out=65,in=25] (P1);

    % P2: leaves 1 and 3.
    \draw[qone]
        (P2) edge[orbitloop,out=185,in=145] (P2);
    \draw[qthree]
        (P2) edge[orbitloop,out=275,in=235] (P2);

    % P3: leaves 1 and 2.
    \draw[qone]
        (P3) edge[orbitloop,out=35,in=-5] (P3);
    \draw[qtwo]
        (P3) edge[orbitloop,out=305,in=265] (P3);

    %--------------------------------------------------
    % Visible basis vertices
    %--------------------------------------------------

    % Triangle K_3.
    \foreach \v in {XXZ,XZX,ZXX,ZZZ} {
        \node[hub,fill=yellow!35] at (c\v) {};
    }

    % Path centred at qubit 1.
    \foreach \v in {YXX,YXZ,YZX,YZZ} {
        \node[hub,fill=orange!30] at (c\v) {};
    }

    % Path centred at qubit 2.
    \foreach \v in {XYX,XYZ,ZYX,ZYZ} {
        \node[hub,fill=cyan!25] at (c\v) {};
    }

    % Path centred at qubit 3.
    \foreach \v in {XXY,XZY,ZXY,ZZY} {
        \node[hub,fill=violet!25] at (c\v) {};
    }

    \foreach \v in \allbases {
        \node[vlabel] at (c\v) {$\v$};
    }

    %--------------------------------------------------
    % Miniature graphs inside the LC-orbit vertices
    %
    % Arguments:
    %   #1 = orbit vertex name
    %   #2 = fill colour
    %   #3 = list of graph edges
    %
    % Qubit positions are fixed in all four graphs:
    %   1 at the top, 2 bottom left, 3 bottom right.
    %--------------------------------------------------

    \newcommand{\draworbitgraph}[3]{%
        \node[hub,minimum size=14mm,fill=#2]
            at (c#1) {};

        \begin{scope}[shift={(c#1)}]

            \coordinate (g#1-1) at (90:\MiniGraphR);
            \coordinate (g#1-2) at (210:\MiniGraphR);
            \coordinate (g#1-3) at (330:\MiniGraphR);

            % Graph edges.
            \foreach \u/\v in {#3} {
                \draw[
                    black,
                    line width=0.9pt,
                    line cap=round
                ]
                    (g#1-\u) -- (g#1-\v);
            }

            % Graph vertices and qubit labels.
            \foreach \v/\a in {1/90,2/210,3/330} {
                \fill[black]
                    (g#1-\v) circle[radius=0.6mm];

                \node[
                    font=\tiny,
                    inner sep=0pt
                ] at (\a:\MiniLabelR) {$\v$};
            }

        \end{scope}
    }

    % Triangle K_3.
    \draworbitgraph{K}{yellow!35}{1/2,1/3,2/3}

    % Path centred at qubit 1.
    \draworbitgraph{P1}{orange!30}{1/2,1/3}

    % Path centred at qubit 2.
    \draworbitgraph{P2}{cyan!25}{1/2,2/3}

    % Path centred at qubit 3.
    \draworbitgraph{P3}{violet!25}{1/3,2/3}

    %     \node[font=\Large] at (0,-5.0)
    %     {$\mathcal{S}(\ket{K_3})$};

    % \node[font=\Large] at (\OrbitX,-5.0)
    %     {$\mathcal{LC}(\ket{K_3})$};

        \node[font=\Large] at (-4.5,3)
        {$\mathcal{S}(\ket{K_3})$};

    \node[font=\Large] at (-3+\OrbitX,3)
        {$\mathcal{LC}(\ket{K_3})$};

\end{tikzpicture}
\end{figure}
\end{example}
\vspace{-9mm}
In the above example, the local Clifford symmetries of $\ket{K_3}$ cause four distinct bases to share each fundamental graph, so that the LC-orbit $\mathcal{LC}(\ket{K_3})$ loses information that the switching graph $\mathcal{S}(\ket{K_3})$ retains. Luckily such symmetries are not generic: we show in theorem \ref{thm:random_graph_no_sym} in the appendix that, with high probability, $k=1$ for random graph states. In this case, the map $B\mapsto G_B$ \emph{is} a bijection from bases to the fundamental graphs of the state. 

\newpage

\subsection{Vertex-minors}
Bases are particularly well-behaved with regard to so-called \emph{vertex-minors}. Vertex-minors are a well-studied subject in both the math~\cite{oum2005rank, oum2023rank, campbell2026erdHos} and the quantum information literature~\cite{dahlberg2018transforming, dahlberg2022complexity, goodenough2026asymptotic, davies2025preparing}. Informally, the vertex-minors of a stabilizer state $\ket{\psi}$ are all the states that can be reached from $\ket{\psi}$ using local operations within the stabilizer formalism.
\begin{definition}[Vertex-minors of stabilizer states~\cite{goodenough2026asymptotic}]
Let $\ket{\psi}$ be a stabilizer state. A state $\ket{\phi}$ is a vertex-minor of $\ket{\psi}$ if $\ket{\phi}$ can be reached from $\ket{\psi}$ using destructive local Pauli measurements and local Clifford unitaries conditioned on those outcomes.
\end{definition}
We note that vertex-minors are commonly expressed in terms of graphs/graph states only; we use the more general notion for stabilizer states, as used in~\cite{goodenough2026asymptotic}.

The bases of vertex-minors of $\ket{\psi}$ are easily expressed in terms of the bases of $\ket{\psi}$. To provide some intuition, we will consider the particularly simple setting where $P$ is contained in at least one basis $B$ of $\ket{\psi}$; we give the general case in the appendix, see theorem~\ref{thm:bases_vm2}.

\begin{theorem}[Bases of vertex-minors~\cite{bouchet1998multimatroids}]\label{thm:bases_vm1}
Let $\mathcal{B}_{\psi}$ be the bases of a stabilizer state
$\ket{\psi}$, and let $P$ be a Pauli string with support
$W\subseteq V$ such that $P$ is contained in at least one basis. Let $\ket{\phi}$ be the stabilizer state on
$V\setminus W$ obtained by measuring each $v\in W$ in the
$P_v$ basis. The bases of $\ket{\phi}$ are given by
\[
\mathcal{B}_{\phi}
=
\left\{
B|_{V\setminus W}
\;\middle|\;
B\in\mathcal{B}_{\psi},
\quad
P\subseteq B
\right\},
\]
where $B|_{V\setminus W}$ denotes the restriction of $B$ to $V\setminus W$.
\end{theorem}
\begin{proof}
Since $P$ is a subset of some basis $B$, the outcomes of measuring $P$ are themselves uniform. A Pauli string $B'$ on the remaining qubits is a basis after the measurement exactly when the joint measurement $P\cup B'$ was completely random before the measurement.
\end{proof}

Here are some examples.

\begin{example}
Recall that the bases of $\ket{00}$ and $\ket{00}+\ket{11}$ are given by
$\lbrace XX, XY, YX, YY\rbrace$ and
$\lbrace XY, XZ, YX, YZ, ZX, ZY\rbrace$, respectively.

First consider measuring the first qubit of $\ket{00}$ in the $X$ basis.
The bases which agree with this measurement are $XX$ and $XY$, and
restricting these to the second qubit gives
\[
\lbrace X,Y\rbrace.
\]
Indeed, after the measurement the second qubit is Pauli-equivalent to the state $\ket{0}$,
whose bases are precisely $\lbrace X,Y\rbrace$.

Similarly, consider measuring the first qubit of
$\ket{00}+\ket{11}$ in the $X$ basis. The bases which agree with this
measurement are $XY$ and $XZ$. Restricting to the unmeasured qubit gives
\[
\lbrace Y,Z\rbrace,
\]
which are precisely the bases of the resulting state $\ket{\pm}$, depending on the measurement outcome.
\end{example}

The above answers a question from~\cite{burchardt2025foliage}, where the authors ask if there is a relation between the LC-orbit of a graph state, and the LC-orbits of its vertex-minors. Instead, theorem \ref{thm:bases_vm1} shows that this connection is most naturally understood in terms of bases. We now express this purely in terms of switching graphs. 

\begin{theorem}
Let $\ket{\psi}$ be a stabilizer state with switching graph $\mathcal{S}(\ket{\psi})$. For $P$ a Pauli string, denote by $\mathcal{S}(\ket{\psi})\,[P]$ the induced subgraph on the set of vertices $B$ such that $P\subseteq B$. Then the set of switching graphs of all vertex-minors is equal to the set of all graphs $\lbrace \mathcal{S}(\ket{\psi})\,[P]\mid \exists B\in \mathcal{B}_\psi~\textrm{s.t.}~P\subseteq B\rbrace $ (up to isomorphism).
\end{theorem}
\begin{proof}
The case where $P\subseteq B$ for some basis follows from theorem~\ref{thm:bases_vm1}. The case when $P$ is not contained in any basis $B$ is dealt with in theorem~\ref{thm:bases_vm2} in the appendix, and does not introduce any new vertex-minors. This last fact follows from the \emph{scum theorem} for multimatroids~\cite{bouchet1998multimatroids, chun2017inductive}.
\end{proof}
\begin{example}
As an example, let us first consider measuring the first qubit of $\ket{K_3}$ in $X$, $Y$ and $Z$. Note that all three measurements satisfy the conditions of theorem \ref{thm:bases_vm1}. In terms of the switching graph, we find the following.
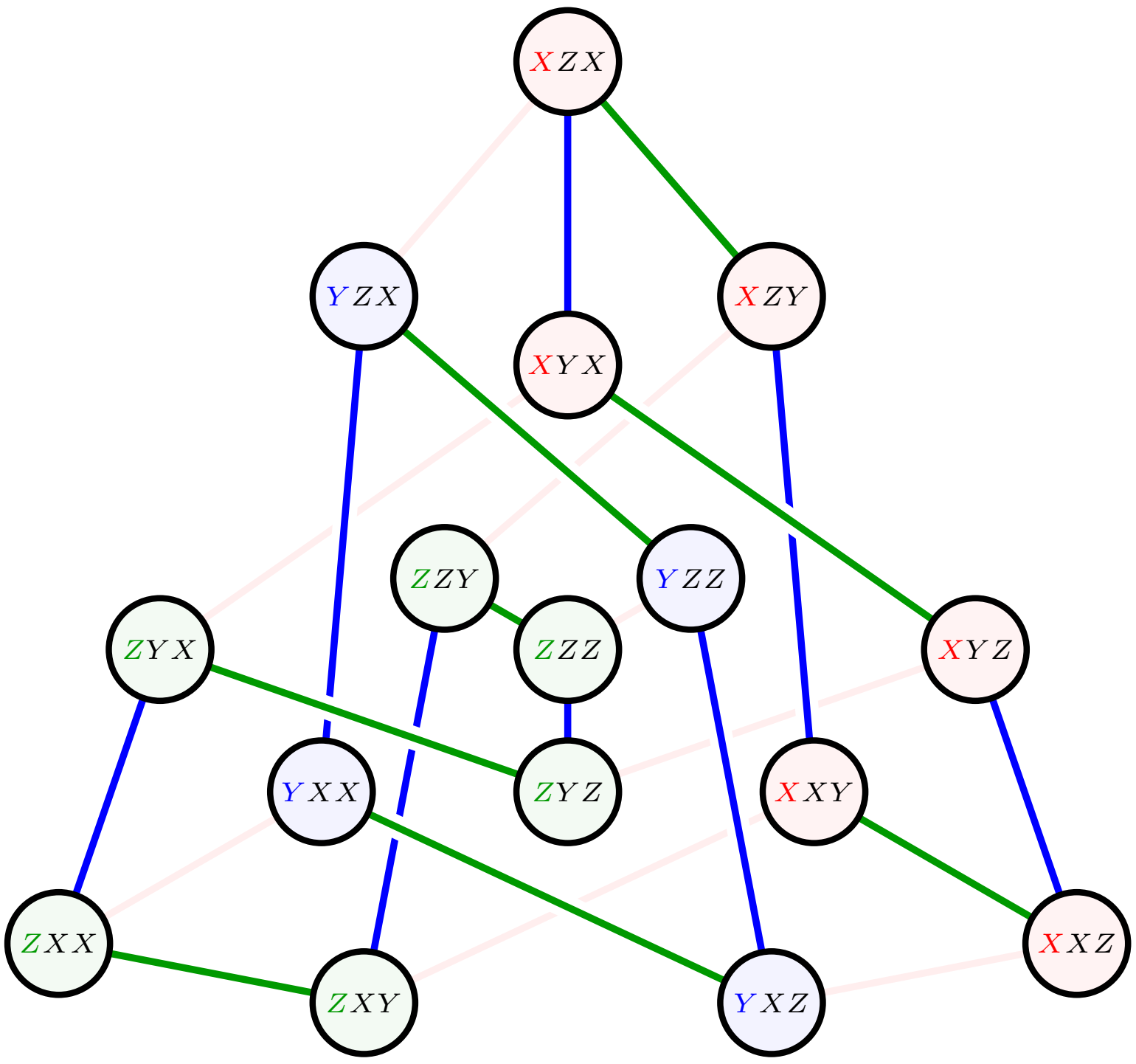
\begin{figure}[H]
    \centering
% Requires: \usepackage{xcolor}
%           \usepackage{tikz}
%           \usetikzlibrary{positioning,arrows.meta}

% Opacity for edges whose endpoints do NOT agree on the first entry.
% 1 = no fading, 0 = invisible.
\newcommand{\firstentryfadeopacity}{0.07}

% Opacity of the vertex fill.
% 1 = solid color, 0 = transparent.
\newcommand{\vertexfillopacity}{0.05}

% Colors associated with the first Pauli entry.
\colorlet{firstXcolor}{red}
\colorlet{firstYcolor}{blue}
\colorlet{firstZcolor}{green!60!black}

\begin{tikzpicture}[
    scale=0.85,
    every node/.style={font=\Large},
    edge/.style={
        line width=1.8pt,
        line cap=round,
        line join=round,
        preaction={draw=white, line width=6pt}
    },
    fadededge/.style={
        edge,
        opacity=\firstentryfadeopacity
    },
    qone/.style={fadededge, draw=red},
    qtwo/.style={edge, draw=blue},
    qthree/.style={edge, draw=green!60!black},
    hubanchor/.style={
        circle,
        minimum size=9.2mm,
        inner sep=0pt
    },
    hub/.style={
        circle,
        draw=black,
        line width=1.6pt,
        minimum size=9.2mm,
        inner sep=0pt,
        text opacity=1
    },
    vlabel/.style={font=\scriptsize}
]

    %--------------------------------------------------
    % Symmetry convention
    %--------------------------------------------------
    % A 120-degree rotation sends
    %
    %     P_1 P_2 P_3  ->  P_2 P_3 P_1.
    %
    % Reflection in the vertical axis sends
    %
    %     P_1 P_2 P_3  ->  P_3 P_2 P_1.
    %
    % Thus the underlying graph has D_3 symmetry.
    % With edge colors, reflection swaps qone <-> qthree
    % and fixes qtwo.

    \def\Rhex{4.30}
    \def\Rout{6.20}
    \def\Rmid{3.00}
    \def\Rin{1.50}

    \newcommand{\placebasis}[3]{%
        \coordinate (c#1) at ({#2*cos(#3)}, {#2*sin(#3)});
    }

    %--------------------------------------------------
    % Coordinates
    %--------------------------------------------------

    % Six-element orbit:
    % {XYZ, YZX, ZXY, ZYX, YXZ, XZY}.
    \placebasis{XYZ}{\Rhex}{0}
    \placebasis{XZY}{\Rhex}{60}
    \placebasis{YZX}{\Rhex}{120}
    \placebasis{ZYX}{\Rhex}{180}
    \placebasis{ZXY}{\Rhex}{240}
    \placebasis{YXZ}{\Rhex}{300}

    % Outer triangle:
    % {XXZ, XZX, ZXX}.
    \placebasis{XZX}{\Rout}{90}
    \placebasis{ZXX}{\Rout}{210}
    \placebasis{XXZ}{\Rout}{330}

    % Middle triangle:
    % {XXY, XYX, YXX}.
    \placebasis{XYX}{\Rmid}{90}
    \placebasis{YXX}{\Rmid}{210}
    \placebasis{XXY}{\Rmid}{330}

    % Inner inverted triangle:
    % {YZZ, ZZY, ZYZ}.
    \placebasis{ZYZ}{\Rin}{270}
    \placebasis{YZZ}{\Rin}{30}
    \placebasis{ZZY}{\Rin}{150}

    % Fixed point.
    \coordinate (cZZZ) at (0,0);

    %--------------------------------------------------
    % Invisible circular anchors
    %--------------------------------------------------
    \def\allbases{%
        XXY,XXZ,XYX,XYZ,%
        XZX,XZY,YXX,YXZ,%
        YZX,YZZ,ZXX,ZXY,%
        ZYX,ZYZ,ZZY,ZZZ%
    }

    \foreach \v in \allbases {
        \node[hubanchor] (\v) at (c\v) {};
    }

    %--------------------------------------------------
    % Exchange edges
    %--------------------------------------------------

    % qubit 1, red
    % These change the first entry, so they are faded.
    \draw[qone] (XXY) -- (ZXY);
    \draw[qone] (XXZ) -- (YXZ);
    \draw[qone] (XYX) -- (ZYX);
    \draw[qone] (XYZ) -- (ZYZ);
    \draw[qone] (XZX) -- (YZX);
    \draw[qone] (XZY) -- (ZZY);
    \draw[qone] (YXX) -- (ZXX);
    \draw[qone] (YZZ) -- (ZZZ);

    % qubit 2, blue
    % These preserve the first entry, so they stay opaque.
    \draw[qtwo] (XXY) -- (XZY);
    \draw[qtwo] (XXZ) -- (XYZ);
    \draw[qtwo] (XYX) -- (XZX);
    \draw[qtwo] (YXX) -- (YZX);
    \draw[qtwo] (YXZ) -- (YZZ);
    \draw[qtwo] (ZXX) -- (ZYX);
    \draw[qtwo] (ZYZ) -- (ZZZ);
    \draw[qtwo] (ZXY) -- (ZZY);

    % qubit 3, green
    % These preserve the first entry, so they stay opaque.
    \draw[qthree] (XXY) -- (XXZ);
    \draw[qthree] (XYX) -- (XYZ);
    \draw[qthree] (XZX) -- (XZY);
    \draw[qthree] (YXX) -- (YXZ);
    \draw[qthree] (YZX) -- (YZZ);
    \draw[qthree] (ZXX) -- (ZXY);
    \draw[qthree] (ZYX) -- (ZYZ);
    \draw[qthree] (ZZY) -- (ZZZ);

    %--------------------------------------------------
    % Visible vertices and labels, drawn last
    %--------------------------------------------------

    % This macro:
    %   - fills the vertex according to its first Pauli,
    %   - uses the same color for the first letter,
    %   - leaves the remaining two letters black.
    \newcommand{\drawbasisvertex}[3]{%
        \def\currentfillcolor{firstZcolor}%
        \if#1X%
            \def\currentfillcolor{firstXcolor}%
        \else\if#1Y%
            \def\currentfillcolor{firstYcolor}%
        \fi\fi
        \node[
            hub,
            fill=\currentfillcolor,
            fill opacity=\vertexfillopacity
        ] at (c#1#2#3) {};
        \node[vlabel] at (c#1#2#3) {%
            $\textcolor{\currentfillcolor}{#1}#2#3$%
        };
    }

    \foreach \v in \allbases {
        \expandafter\drawbasisvertex\v
    }

\end{tikzpicture}
    \caption{Switching graphs of the vertex-minors when measuring the first qubit in $X$, $Y$ or $Z$.}
    \label{fig:basis_exchange_stabilizers_minors}
\end{figure}
Here we have highlighted the three subgraphs of the switching graph $\mathcal{S}(\ket{K_3})$ whose vertices are the bases with their first entry equal to $X$, $Y$ and $Z$, respectively. The subgraph corresponding to $X$ (red vertices) forms a six-cycle, and we recognize this as the switching graph of a two-qubit entangled stabilizer state, see example \ref{example:basis_graphs}. In other words, measuring the first qubit of $\ket{K_3}$ in $X$ yields (up to rotations) a Bell pair on the remaining two qubits. Something similar holds for measuring in $Z$: the subgraph of bases with first entry equal to $Z$ forms a six-cycle as well (green vertices), so measuring the first qubit in $Z$ will yield an entangled state. Conversely, measuring the first qubit in $Y$ will leave the remaining two qubits unentangled, since the subgraph corresponding to $Y$ (blue vertices) is a four-cycle, i.e., the switching graph of a separable two-qubit state.
\end{example}
Last but not least, bases serve as a natural introduction to \emph{multimatroids}.

\newpage
\section{Multimatroids}\label{sec:multimatroids}
In this paper the bases $\mathcal{B}$ were the main actors. These are the local Pauli measurements on all qubits such that each of the $2^n$ outcomes is equally likely. Naively such measurements would seem useless to understand entanglement. We have seen, however, that bases suitably generalize the notion of graphs, and that switchings generalize the notion of local complementations. 

Instead of considering only bases (which yield $2^n$ outcomes at random), we can ask for each Pauli string $P$ how many outcomes occur. Each such measurement will have $2^r$ outcomes, all equally likely, for some integer $r$. The integer $r$ we call the \emph{rank} of $P$, and quantifies the randomness observed when measuring according to $P$~\cite{goodenough2026asymptotic}. Bases are then exactly the Pauli strings with the maximum rank of $n$.

A multimatroid\footnote{This informal definition holds for the very specific context relevant to qubit stabilizer states. Multimatroids are, by themselves, more general.} can then be defined by the function that assigns to each Pauli string $P$ its rank (with respect to a given stabilizer state)~\cite{bouchet1997multimatroids, bouchet1998multimatroids, bouchet2001multimatroids, brijder2015isotropic1, brijder2018orienting, goodenough2026asymptotic}. The rank function determines a great deal of the measurement statistics when performing local Pauli measurements~\cite{goodenough2026asymptotic}, and suitably generalizes the notion of a rank function from matroid theory~\cite{oxley2006matroid}. While we will not define matroids here, it suffices to say that the notions of bases and circuits discussed in this work are natural multimatroid generalizations of bases and circuits in matroid theory. Similarly, the construction of $G_B$, which seems rather strange at first sight, is a direct analogue of the well-studied notion of fundamental graphs from matroid theory~\cite{krogdahl1977dependence, las1980fundamental}.

Rank functions, bases and circuits all provide ways to define (multi)matroids, but there are many more. Another example is given by the \emph{cycles} of a (multi)matroid, which are the analogue of the more familiar notion of stabilizers in the quantum community. The fact that there are many definitions of (multi)matroids that are not obviously equivalent allows one to use the perspective most effective for the problem at hand. These equivalent definitions are endearingly referred to as \emph{cryptomorphisms} in the matroid community. The follow-up work~\cite{goodenough2026stabilizer} is devoted to further establishing these deep connections between stabilizer states and multimatroid theory.

\section*{Acknowledgements}
I am grateful for discussions with Robert Brijder, Nathan Claudet, Rose McCarty, Axel Dahlberg, Gecia Bravo-Hermsdorff, Lee Gunderson, Isma\"el Septembre, Paul Gunnells, and Steven Noble (in no particular order). I am in particular indebted to Fynn Otto, Frederik Hahn, Nathan Claudet, Adam Fredriksson and Rapha\"el Mothe for feedback on the writing.

I acknowledge support from the Alexander von Humboldt Foundation. I used ChatGPT to clean up my writing, to find a proof of theorem \ref{thm:diameter} using only notions introduced here, to fully prove theorem~\ref{thm:random_graph_no_sym}, to create all figures in TikZ, and to create the code that visualizes basis switchings and fundamental circuits in~\cite{visualization}.

\newpage

\bibliographystyle{plain}
\bibliography{references}

\appendix

\section{Proofs}

\begin{lemma}\label{lemma:extend_to_basis2}
Let $\ket{\psi}$ be a stabilizer state, and $Q$ an arbitrary (not necessarily full-weight) Pauli string such that measuring according to $Q$ yields $2^{\left|\textrm{supp}(Q)\right|}$ random outcomes. Then there exists a basis $B\in \mathcal{B}_\psi$ such that $Q\subseteq B$.
\end{lemma}
\begin{proof}
If $Q$ is full-weight we are done, since then $Q$ is a basis. So suppose that $Q$ is not full-weight. Consider an arbitrary qubit $w$ not in the support of $Q$, and let $P_w$ be an arbitrary non-identity Pauli on $w$. Consider measuring according to $Q\cup P_w$. Since $Q$ yields random outcomes, $Q\cup P_w$ will yield non-random outcomes iff measuring according to $Q$ leaves qubit $w$ in an eigenstate of $P_w$ (in particular, random outcomes are obtained independent of $P_w$ if qubit $w$ is entangled with any other qubit). Since $Q$ is fixed, we can choose $P_w$ such that this never happens by using property~\ref{property:stab_no_signs_meas}.

This implies that $P_w$ can be chosen such that $Q\cup P_w$ yields random outcomes, similar to the original $Q$. Iterating the above procedure yields a full-weight Pauli string $B$ such that measuring according to $B$ yields all $2^n$ outcomes with equal probability (so that $B\in \mathcal{B}_{\psi}$), and $Q\subseteq B$. This proves the claim.
\end{proof}

\medskip

\begin{theorem}[Upper bound on number of local complementations~\cite{bouchet1990kappa}]\label{thm:diameter2}
If $G$ and $H$ are locally equivalent graphs on $n$ vertices, then at most $\frac{3}{2}n$ local complementations are needed to transform $G$ into $H$.
\end{theorem}
\begin{proof}
Rephrasing this in terms of bases, we aim to show that the number of switchings needed to go from one basis $B$ to any other basis $B'$ is bounded by $\frac{3}{2}n$. Part of our proof follows Bouchet's approach from~\cite{bouchet1990kappa} (albeit with slightly different notation). We prove the result by inducting on the number of entries $d$ in which $B$ and $B'$ differ. We first show that the induction step holds when $B$ and $B'$ have an entry $v$ such that $\left(B\star v\right)_v=B'_v$; this is the easy case. We then show that the induction also holds when switching $B$ at $v$ does not map that entry to $B'_v$

Let $D(B, B'):=\lbrace v\in V \mid  B_v\neq B_v'\rbrace$ be the entries in which $B$ and $B'$ differ. Let $d:=\left|D(B, B')\right|$, and let $s(B, B')$ be the smallest number of switchings needed to go from $B$ to $B'$. We will prove the slightly stronger claim that $s(B, B')\leq \frac{3}{2}d$, from which $s(B, B')\leq \frac{3}{2}n$ follows directly.

We prove the claim by induction on $d$. Clearly if $d=0$, then $s(B, B')\leq \frac{3}{2}d$, so suppose $d>0$.

If there exists a $v\in D(B, B')$ such that $\left(B\star v\right)_v=B_v'$, then $\left|D(B\star v, B')\right|=d-1$. By induction, 

\[s(B, B')\leq s(B\star v, B')+1\leq \frac{3}{2}\left|D(B\star v, B')\right|+1\leq \frac{3}{2}\left|D(B, B')\right|.\]

We may thus assume that $(B\star v)_v\neq B'_v$ for every $v\in D(B, B')$. 

Choose any $v\in D(B, B')$, and consider the fundamental circuit $g_v^B$. By definition, $g_v^B$ is the unique circuit in $B\cup \omega_v$. Note that $(g_v^B)_v$ is equal to $B_v'$, since changing the $v$'th entry of $B$ to $(B\star v)_v$ yields a basis, which cannot contain a stabilizer by definition~\ref{def:bases}. Together with our assumption that $(B\star v)_v \neq B_v'$, we thus have that $g_v^B$ is contained in $B\cup \lbrace B_v'\rbrace$, and thus $\left(g_v^B\right)_v=B_v'$. Now $g_v^B$ cannot be contained in $B'$ since $B'$ is a basis, yet $(g_v^B)_v=B_v'$ by the previous argument. This must then mean that there exists an entry $w$ different from $v$ such that $w\in \textrm{supp}(g_v^B)$ and that $(g_v^B)_w\neq B_w'$, since otherwise $B'$ would contain the stabilizer $g_v^B$. Since $(g_v^B)_w=B_w$, we have that both $v, w$ are in $D(B, B')$. More strongly, $w\in D(B, B')\cap \textrm{supp}\left(g_v^B\right)$, i.e.,~$v$ and $w$ are adjacent in $G_B$.

For such pairs $v, w$, the \emph{pivot} on $B$ at $\lbrace{v,w\rbrace}$ is defined as
\[B\wedge vw:= B\star v \star w \star v\ ,\]
where we dropped the set notation~\cite{oum2005rank, bouchet2001multimatroids}. This is motivated by the fact that the order does not matter for pivots, i.e.~$B\wedge vw = B\wedge wv$~\cite{oum2023rank}.

We now show that $\left(B\wedge vw\right)_v=B_v'$ and $\left(B\wedge vw\right)_w=B_w'$, i.e.~a pivot maps both the $v$ and $w$ entries of $B$ to $B_v'$ and $B_w'$, respectively. Let $K=B\setminus \lbrace B_v, B_w\rbrace$. Pivoting on $vw$ then gives

\begin{align}
B = &\,K\cup \lbrace B_v,\,B_w\rbrace \nonumber\\
\xrightarrow{\star v} &\,K\cup \lbrace (B\star v)_v,\, B_w\rbrace\nonumber\\
\xrightarrow{\star w} &\,K\cup \lbrace (B\star v)_v,\, B_w'\rbrace \nonumber\\
\xrightarrow{\star v} &\,K\cup \lbrace B_v',\, B_w'\rbrace\ , \nonumber
\end{align}
as desired since both the $v$ and $w$ entries now match $B'$ (without changing any of the other entries). The above follows from the fact that $(B\star v \star w)_w\neq (B\star w)_w \iff v\sim_{G_B}w$, which is implicit from theorem 7.6 of~\cite{bouchet1988graphic} and theorem 13.4 of~\cite{kim2024vertex}.

Thus, by performing three switchings, the entries $v, w$ of $B$ and $B'$ can be made to match. By the induction hypothesis, we have that
\begin{align}
s(B,B') &\leq 3+s(B\wedge vw,B')\nonumber \\
&\leq 3+\frac{3}{2}\left|D(B\wedge vw,B')\right|\nonumber \\
&=3+\frac{3}{2}(d-2)\nonumber \\
&=\frac{3}{2}d\ . \nonumber
\end{align}
Together with the case where a single switching decreases $d$, this proves that
\[
s(B,B')\leq \frac{3}{2}d\leq \frac{3}{2}n\, , 
\]
as desired.
\end{proof}

\medskip

In the below we define the index $k$ as the cardinality of the preimage of the map $B\mapsto G_B$, see~\cite{bouchet1993recognizing}, which we also show to be well-defined.

\begin{theorem}\label{thm:k_is_sym}
The index $k$ of a stabilizer state $\ket{\psi}$ is equal to the cardinality of $\mathcal{C}_\psi$, where $\mathcal{C}_\psi$ is the group of local Clifford symmetries of $\ket{\psi}$ up to Paulis.
\end{theorem}
\begin{proof}
It suffices to show that 1) two bases $B, B'$ map to the same fundamental graph (i.e.~$G_B=G_{B'}$) if and only if $CBC^\dagger=B'$ for some $C\in \mathcal{C}_\psi$, and that 2) the action of the group $\mathcal{C}_\psi$ on the set of bases $\mathcal{B}_\psi$ by conjugation is \emph{free}, i.e.~for every pair $B, B'$, there exists at most one $C\in \mathcal{C}_\psi$ such that $CBC^\dagger = B'$. The first condition implies that the preimage of a fundamental graph $G_B$ is precisely the orbit of $B$ under the action of $\mathcal{C}_\psi$. The second condition then implies that the size of this orbit is equal to the size of $\mathcal{C}_\psi$.

\medskip
We show now the first condition. We first consider the forward direction. 
Let $B, B'$ be such that $G_B=G_{B'}$. By the proof of theorem~\ref{theorem:bases_surj_into_graphs}, we have that there exist local Cliffords $C_B$ and $C_{B'}$ such that $\left(C_B\right)B\left(C_B\right)^\dagger=Z^{\otimes n}$, and a similar statement for $B'$ and $C_{B'}$. Theorem~\ref{theorem:bases_surj_into_graphs} then also implies that $C_B\ket{\psi}=\ket{G_B}=\ket{G_{B'}}=C_{B'}\ket{\psi}$, so that $\left(C_{B'}\right)^\dagger C_B$ is a symmetry of $\ket{\psi}$ (modulo Paulis), which maps $B$ to $B'$ under conjugation.

For the other direction, note that if $C\in \mathcal{C}_\psi$ and $C$ maps $B$ to $B'$, then applying $C$ to the state and to $B$ preserves the supports of the fundamental circuits. This implies that $G_B=G_{B'}$.
\medskip

Now we show that this action is free. We use the fact that $\mathcal{C}_\psi$ acts freely on the set of bases iff $CBC^\dagger =B$ implies that $C$ is the identity. So assume that $C$ fixes some $B$. $C$ then necessarily fixes $B\cup \omega_v$, and, by definition~\ref{def:fund_circuits}, fixes $g_v^B$ for every qubit $v$. But since $\left(g_v^B\right)_v$ and $B_v$ are two distinct Paulis that are fixed by $C$, $C$ needs to fix all elements in $\omega_v$ for each $v$, proving the claim.
\end{proof}

Using the characterization of the index as the cardinality of the group of local Clifford symmetries, we can now show that such local Clifford symmetries are rare, as made precise by the following theorem.
\begin{theorem}\label{thm:random_graph_no_sym}
With high probability as $n$ tends to infinity, the index $k(G)$ of a uniformly random graph $G\sim G_{n, 1/2}$ equals $1$. Equivalently, with high probability a random graph state has no non-trivial local Clifford symmetries (that are not Pauli symmetries) as $n$ tends to infinity.
\end{theorem}
\begin{proof}
We sketch now the proof idea. We show that, with high probability, the canonical graph state generators $\lbrace g_v\rbrace_{v\in V}$ are the only stabilizers of $\ket{G}$ with support equal to the closed neighborhoods $\lbrace \overline{N}_v\rbrace_{v\in V}$. To make the above statement more plausible before proving it (which we do in a different manner than the argument we show now), note that the normalized weight distribution of stabilizers of a random graph state is well-approximated by a binomial distribution $\mathbb{B}(n, p=\frac{3}{4})$~\cite{miller2023shor}. Now each generator $g_v$ has weight approximately $n/2$, which can be checked to have exponentially small probability when sampled from $\mathbb{B}(n, p=\frac{3}{4})$. Having made this step more plausible, let us assume that the supports of the generators $g_v$ are indeed unique. Then since local Clifford unitaries cannot change the support of stabilizers, we have that every putative local Clifford symmetry maps every canonical graph state generator to itself. In particular, if there are no isolated vertices then there always exists a $w$ that is a neighbor of $v$, such that both $\left(g_v\right)_v=X_v$ and $\left(g_w\right)_v=Z_v$ are fixed under the local Clifford symmetry. This implies that the local Clifford symmetry is trivial.

We now show that $g_v$ is the only stabilizer with support $\overline{N}_v$ with high probability. First, the case where $G$ has any isolated vertex happens with exponentially small probability, so we do not need to consider it. So let us assume that the graph has no isolated vertices. We use that every stabilizer can be written uniquely as

\begin{align}
\Gamma_W := \prod_{w\in W}g_w=  \prod_{w\in W}X_wZ_{N_w}, \nonumber
\end{align}
where $W\subseteq V$ is an arbitrary subset of vertices. The stabilizer $\Gamma_W$ has support $W\cup \textrm{Odd}(W)$, where $\textrm{Odd}(W)$ is the set of vertices in $V$ that have an odd number of neighbors in $W$.

Fix a vertex $v$. For the support of another stabilizer $\Gamma_W$ to agree with the support $\textrm{supp}\left(g_v\right)=\overline{N}_v$ of $g_v$, we necessarily require that $W$ satisfies $\emptyset \neq W\subseteq \overline{N}_v$, and that $W\neq \lbrace v\rbrace$.

Let us condition on $N_v=D$, and set $d=|D|$. All edges not incident
to $v$ remain independent, each being present with probability
$1/2$. The standard terminology is that the edges we have not conditioned on/assumed anything about are \emph{unrevealed}. 
There are two cases to consider for $W$, either $v\not\in W$, or $v \in W$ so that we can write $\Gamma_{W}=g_v \Gamma_T$, where $T=W\setminus \lbrace v\rbrace$.

Fix such a set $T$, and write $t=|T|$. For each vertex
$u\notin T\cup\{v\}$, $\Gamma_T$ has a $Z$ at $u$ iff $u\in \textrm{Odd}(T)$, iff the parity of $|N_u\cap T|$ is $1$. But note that $|N_u\cap T|$ is a uniformly
random bit. In fact, these bits are independent for distinct vertices
$u, w\notin T\cup\lbrace v\rbrace$, since they depend on disjoint sets of unrevealed edges.

Consider first the case $v\notin W$, so that $W=T$.
The stabilizer $\Gamma_T$ already acts nontrivially on every
qubit in $T$. To have support precisely $D\cup\{v\}$, $\Gamma_T$ needs to have a $Z$
on every qubit in $D\setminus T$, and $I$ on every qubit
outside $D\cup\{v\}$. This thus fixes the parity of
$|N_u\cap T|$ at each of the $n-1-t$ vertices in
$u\notin T\cup\{v\}$.
The additional requirement at $v$ is that $t$ be odd, since
every vertex of $T$ is adjacent to $v$. Ignoring this additional
requirement gives
\begin{align}
\textrm{Prob}\left[
\textrm{supp}(\Gamma_T)=D\cup\{v\}
\mid N_v=D
\right]
\leq 2^{-(n-1-t)},\nonumber
\end{align}
which is independent from $d$.

Consider next the case $v\in W$, so that we can write $W=T\cup\{v\}$.
The stabilizer $g_v\Gamma_T$ already acts nontrivially on every
qubit in $T\cup\{v\}$. For its support to equal $D\cup\{v\}$,
we require $\Gamma_T$ to act trivially on every qubit
$u\notin T\cup\{v\}$. Indeed, a $Z$ in $\Gamma_T$ would
either cancel the $Z$ in $g_v$ (if $u\in D$) or introduce
a $Z$ where $g_v$ acts trivially (if $u\notin D$).
Equivalently, $|N_u\cap T|$ must be even for every
$u\notin T\cup\{v\}$.
Equivalently, $|N_u\cap T|$ must be even for every
$u\notin T\cup\{v\}$. These are again $n-1-t$ independent
parity requirements, and hence
\begin{align}
\textrm{Prob}\left[
\textrm{supp}(\Gamma_{T\cup \lbrace v\rbrace})=D\cup\{v\}
\mid N_v=D
\right]
\leq 2^{-(n-1-t)},\nonumber
\end{align}

Let $B_v$ denote the event that some stabilizer other than
$g_v$ has support $\overline{N}_v$. Taking a union bound over
the two cases where $v\in W$ and $v\not\in W$, and all nonempty subsets $T\subseteq D$ yields
\begin{align}
\textrm{Prob}(B_v\mid N_v=D)
&\leq
2\sum_{t=1}^{d}\binom{d}{t}2^{-(n-1-t)}\nonumber \\
&= \frac{4(3^d-1)}{2^n}.
\label{eq:gen_supp_collision}
\end{align}

It remains to account for the randomness of $N_v$.
Note that the size $d$ of the neighborhood of $v$ follows a binomial distribution, i.e.~$d\sim\operatorname{Bin}(n-1,1/2)$. A standard tail bound then yields
\begin{align}
\textrm{Prob}\left[d(v)>\frac{3}{5}(n-1)\right]
\leq e^{-(n-1)/50}\label{eq:app_of_constant},
\end{align}
where the choice of constant $\frac{3}{5}$ will be shown to be convenient in not too long.

We now show the claim for every small enough neighborhood $D$, i.e.~$d\leq 3(n-1)/5$. Equation~\eqref{eq:gen_supp_collision} gives
\begin{align}
\textrm{Prob}(B_v\mid N_v=D)
\le
2\left(\frac{3^{3/5}}{2}\right)^{n-1}.\nonumber
\end{align}
Thus, averaging over $N_v$ and taking a union bound over
all vertices,
\begin{align}
\textrm{Prob} \left[\bigcup_{v\in V}B_v\right]
\le
n e^{-(n-1)/50}
+
2n\left(\frac{3^{3/5}}{2}\right)^{n-1} \label{eq:prob_bound_single_v}\ ,
\end{align}
which goes to zero as $n$ tends to infinity since $\frac{3^{3/5}}{2}<1$, motivating the choice of constant in Eq.~\eqref{eq:app_of_constant}.

Finally, the above analysis was for a single vertex $v$. Repeating the same argument for all $n$ vertices and using a union bound gives the desired result, since Eq.~\eqref{eq:prob_bound_single_v} decays exponentially fast in $n$.
\end{proof}

\medskip

Below we provide the fully general scenario of how the set of bases changes when performing local Pauli measurements according to some Pauli string $P$. In particular, we consider the case where there does not need to exist a basis $B$ that contains $P$. The following theorem covers that case.

\begin{theorem}[Bases of vertex-minors once more, \cite{bouchet1998multimatroids}]\label{thm:bases_vm2}
Let $\mathcal{B}_{\psi}$ be the bases of a stabilizer state
$\ket{\psi}$, and let $P$ be an arbitrary Pauli string with support equal to 
$W\subseteq V$. Let $\ket{\phi}$ be the stabilizer state on
$V\setminus W$ obtained by measuring each $v\in W$ in the
$P_v$ basis. The bases of $\ket{\phi}$ are given by
\[ \mathcal{B}_{\phi}
=
\left\{
B|_{V\setminus W}
\mid B\in\mathcal{B}_{\psi}, \quad |B\cap P| =r
\right\},
\]
where $r$ is the \emph{rank of} $P$, i.e.~the integer $r$ such that $2^r$ is the number of possible outcomes when measuring according to $P$, see section \ref{sec:multimatroids}.
\end{theorem}
\begin{proof}
We briefly recall the proof for the case where $P$ is contained in at least one basis, say $B$. Since $B$ is a basis, measuring any subset of qubits according to $B$ yields uniformly random outcomes as well. Thus measuring according to $P$ yields random outcomes. A Pauli string $B'$ on the unmeasured qubits is a basis for the post-measurement state iff $P\cup B'$ would have yielded completely random outcomes on the original state, i.e.~$P\cup B'$ is a basis of the initial state.

Now consider the case where $P$ need not be contained in a basis. Let $B$ be any basis such that $\left|B\cap P\right|=r$. Let $R=\textrm{supp}\left(B\cap P\right)$, and let $P|_{R}$ be the restriction of $P$ to $R$, i.e.~$\left(P|_R\right)_v=P_v$ for $v\in R$, and $\left(P|_R\right)_v=I_v$ for $v\not \in R$. Note that $P|_R$ is contained in $B$, and thus measuring according to $P|_R$ yields $2^{\left|\textrm{supp}\left( P|_R \right)\right|}=2^{r}$ random outcomes. Let $Q=B|_W$, which is clearly contained in the basis $B$. As such, theorem \ref{thm:bases_vm1} applies to $Q$, so that when measuring according to $Q$, the post-measurement state has $B|_{V\setminus W}$ as a basis.

We now claim that the post-measurement state when measuring according to $P$ or $Q$ are the same. This is because the measurements $P_i$ for $i\in  \textrm{supp}(P)
\setminus R$ yield deterministic outcomes, conditioned on the measurement outcomes in $R$.
This follows from the fact that both measuring according to $P$ and $P|_{R}$ yield $2^r$ outcomes uniformly at random.
This in particular implies that after measuring according to $P|_R$, each qubit in $i\in  W
\setminus R$ needs to be a single-qubit stabilizer state, i.e.~not entangled with anything else. Since measuring single-qubit stabilizer states cannot change the state of the other qubits, and our statement was independent of the choice of $B$, we have that the post-measurement states when measuring according to either $P$ or $Q$ are the same. Thus, since $B|_{V\setminus W}$ is a basis of the post-measurement state when measuring according to $Q$, $B|_{V\setminus W}$ is also a basis for the post-measurement state when measuring according to $P$. This holds for any $B$ such that $\left|B\cap P\right|=r$.

We now show that every basis $B'$ of $\ket{\phi}$ arises in the above fashion. To this end, note that by the above argument we can restrict to the case where we measure $Q$ instead of $P$. Since $Q$ is contained in a basis, theorem \ref{thm:bases_vm1} applies, i.e.~$B'$ is a basis of $\ket{\phi}$ iff $B'\cup Q=B$ is a basis of $\ket{\psi}$. Moreover, since $B'$ has support disjoint from $P$, we have $|B\cap P|=|Q\cap P|=r$, and
$B|_{V\setminus W}=B'$. This proves the
converse inclusion, completing the proof.
\end{proof}

Here is a simple example of the more general case of vertex-minors.

\begin{example}
Consider measuring the first qubit
of $\ket{00}$ in the $Z$ basis. None of the bases
\[
\lbrace XX,XY,YX,YY\rbrace
\]
agree with the measurement on the first qubit. Hence the maximal possible
agreement with $P=Z$ is zero, so in this case we retain all four bases.
Restricting them to the unmeasured qubit gives
\[
\lbrace X,Y\rbrace,
\]
which are again precisely the bases of the remaining state $\ket{0}$.
\end{example}

Here we show that, when performing Pauli measurements on a stabilizer state, the post-measurement states corresponding to different outcomes are always equivalent up to Pauli unitaries.

\begin{property}[Stabilizer states are well-behaved under Pauli measurements]\label{property:stab_no_signs_meas2}
Let $\ket{\psi}$ be any stabilizer state, and perform any fixed sequence of Pauli measurements on $\ket{\psi}$. Let $\ket{\phi, O}$ and $\ket{\phi, O'}$ be any two post-measurement states after this process, where $O$ and $O'$ denote two possible sets of measurement outcomes. Then $\ket{\phi, O}$ and $\ket{\phi, O'}$ are equivalent up to Pauli unitaries; in particular, $\ket{\phi, O}$ is an eigenstate of a Pauli operator $Q$ $\iff$ $\ket{\phi, O'}$ is an eigenstate of $Q$.
\end{property}
\begin{proof}
We use the standard update rule for stabilizer states from~\cite{aaronson2004improved}. That is, if $P$ is a stabilizer, the measurement outcome is deterministic and there is only one post-measurement state, so there is nothing left to prove.

If $P$ is not a stabilizer, find a generating set $\lbrace g_1, g_2, \ldots, g_n\rbrace$ of the stabilizer group of the initial state $\ket{\psi}$, such that $g_1$ anti-commutes with $P$ and commutes with all other $g_i$. This is always possible, since if there exist two distinct generators $g_i, g_j$ that anti-commute with $P$, then $g_j$ can be replaced with $g_i\cdot g_j$, which preserves the stabilizer group, while ensuring that $g_i\cdot g_j$ commutes with $P$. The two post-measurement states then have stabilizer groups $\langle \pm P, g_2, \ldots, g_n\rangle$, which are Pauli equivalent since there always exists a Pauli $Q$ that commutes with every generator besides $g_1$~\cite{aaronson2004improved}.

The argument extends to two Pauli measurements, since the choice of $Q$ did not depend on the phases in front of the stabilizers. That is, after performing a second measurement $P'$, the associated Pauli correction $Q'$ is independent of which outcome was observed for the first measurement. This can then be extended to any number of Pauli measurements.
\end{proof}

\end{document}